\documentclass[journal=acsel,manuscript=article]{achemso}

\usepackage{pdfpages}
\usepackage[version=3]{mhchem} %
\usepackage{miller}
\usepackage{multirow}
\usepackage{color}
\usepackage{textgreek}

\graphicspath{{Figures/}}
\usepackage[colorinlistoftodos]{todonotes}

\author{Zeeshan Ahmad}
\affiliation{Pritzker School of Molecular Engineering, University of Chicago, Chicago, Illinois 60637, United States}
\author{Rebecca A. Scheidt}
\affiliation{Chemistry \& Nanoscience Center, National Renewable Energy Laboratory, Golden, Colorado 80401, United States}
\author{Matthew P. Hautzinger}
\affiliation{Chemistry \& Nanoscience Center, National Renewable Energy Laboratory, Golden, Colorado 80401, United States}
\author{Kai Zhu}
\affiliation{Chemistry \& Nanoscience Center, National Renewable Energy Laboratory, Golden, Colorado 80401, United States}
\author{Matthew C. Beard}
\affiliation{Chemistry \& Nanoscience Center, National Renewable Energy Laboratory, Golden, Colorado 80401, United States}
\author{Giulia Galli}
\affiliation{Pritzker School of Molecular Engineering, University of Chicago, Chicago, Illinois 60637, United States}
\alsoaffiliation{Department of Chemistry, University of Chicago, Chicago, Illinois 60637, United States}
\alsoaffiliation{Argonne National Laboratory, Lemont, Illinois 60439, USA}
\email{gagalli@uchicago.edu}

\title{Understanding the Effect of Lead Iodide Excess on the Performance of Methylammonium Lead Iodide Perovskite Solar Cells}


\begin{document}

\begin{tocentry}

\begin{center}
\includegraphics[width=3in]{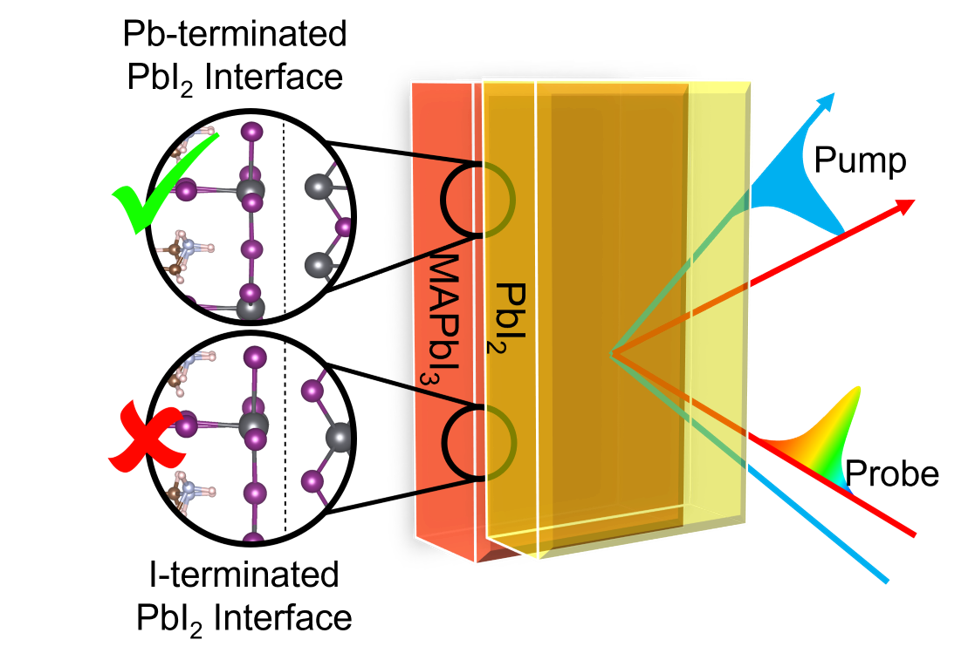}
\end{center}
\end{tocentry}

\begin{abstract}
The presence of unreacted lead iodide in organic-inorganic lead halide perovskite solar cells is widely correlated with an increase in power conversion efficiency.  We investigate the mechanism for this increase by identifying the role of  surfaces and interfaces present between   methylammonium lead iodide perovskite films and excess lead iodide. We show how  type I and II band alignments arising under different conditions result in either passivation of surface defects or hole injection. Through first-principles simulations of solid-solid interfaces, we find that lead iodide captures holes from methylammonium lead iodide and modulates the formation of defects in the perovskite, affecting recombination.   Using surface-sensitive optical spectroscopy techniques, such as transient reflectance and time-resolved photoluminescence, we show how excess lead iodide affects the diffusion and surface recombination velocity of charge carriers in methylammonium lead iodide films.  Our coupled experimental and theoretical results elucidate the role of excess lead iodide in perovskite solar cells.
\end{abstract}

Hybrid organic-inorganic perovskites are ideal candidates for many photovoltaic and optoelectronic applications due to their low-cost solution processability, scalable manufacturing, high absorption coefficients, and long  carrier lifetimes~\cite{Green2014emergence,huangUnderstandingPhysicalProperties2017,yooEfficientPerovskiteSolar2021,liChemicallyDiverseMultifunctional2017}. Perovskite solar cells (PSC) have shown outstanding improvements at the device level over a short period of time with power conversion efficiencies exceeding 25\%~\cite{NRELChart}. The pace of understanding the microscopic origin of such high efficiency, however, has been much slower.

Recently, there has been great interest in improving the efficiency of solution-processed PSCs through the incorporation of excess \ce{PbI2} in the precursor solution~\cite{jacobssonUnreactedPbI2DoubleEdged2016,chenControllableSelfInducedPassivation2014,supasaiFormationPassivatingCH3NH3PbI32013, rooseCriticalAssessmentUse2020, roldan-carmonaHighEfficiencyMethylammonium2015}.  %
Various studies have attempted to understand the influence of the excess \ce{PbI2}, especially as it relates to methylammonium lead iodide (MAPI) which is widely used as a case study to make inferences about other, more complex perovskite compositions. Many of these investigations have primarily focused on the effect of \ce{PbI2} on defect passivation to explain the increase in solar cell efficiency~\cite{jacobssonUnreactedPbI2DoubleEdged2016,barbeLocalizedEffectPbI22018}, including   studies  showing an increase of open-circuit potential when excess \ce{PbI2} is added~\cite{merdasaImpactExcessLead2019,jiangObservationLowerDefect2019}; other studies however observed  no change upon addition of \ce{PbI2} or even a decrease in open-circuit  potential~\cite{jacobssonUnreactedPbI2DoubleEdged2016,rooseCriticalAssessmentUse2020}.
Discrepancies also exist regarding the energy level alignment between MAPI and \ce{PbI2} and its role in determining the effect of \ce{PbI2} on PSCs. ~\citealt{chenControllableSelfInducedPassivation2014} found type I alignment between MAPI and \ce{PbI2} and proposed that \ce{PbI2} increases the power conversion efficiency through passivation at grain boundaries and interfaces, thus  resulting in reduced carrier recombination. However, ~\citealt{calloniStabilityOrganicCations2015} showed that a \ce{PbI2} layer is formed \textit{in situ} during annealing and sputtering at the surface of MAPI and exhibited a type II band alignment which can be used for hole extraction and thus increasing charge separation. Similarly, ~\citealt{rooseCriticalAssessmentUse2020} showed that excess \ce{PbI2} forms a thin ($<5$ nm) layer at the surface of the perovskite that assists in charge extraction. Further,  the MAPI crystallite size has been found to increase with excess \ce{PbI2} in one study~\cite{roldan-carmonaHighEfficiencyMethylammonium2015} while another study  found no significant effect of thickness~\cite{rooseCriticalAssessmentUse2020}. 
\ce{PbI2} is photoactive and a source of  parasitic losses, as it can absorb photons with wavelengths close to 450 nm. However spatially resolved photoluminescence indicated that the  carriers generated  in \ce{PbI2}  migrate to the interface with MAPI and eventually contribute to the open-circuit voltage~\cite{merdasaImpactExcessLead2019}. Detrimental effects of excess \ce{PbI2} have also been observed, namely a worsening of the stability of the \ce{PbI2}-rich  perovskite samples and the presence of a high trap density responsible for lowering  the photovoltaic performance~\cite{gujarRolePbICH2018, wangAdverseEffectsExcess2017, liuExcessPbI2Beneficial2016}.

The wide discrepancies found in the literature suggest that the effects of unreacted \ce{PbI2} in PSCs are still not fully understood and are dependent not only on the amount of excess \ce{PbI2} but also on the processing conditions.  
A microscopic  modeling of  perovskite/\ce{PbI2} interfaces is required to validate experimental findings~\cite{merdasaImpactExcessLead2019,j.dahlmanInterfacesMetalHalide2021,niResolvingSpatialEnergetic2020} and provide detailed  mechanistic explanations, which in turn may be used to derive design rules for device optimization. Previous  studies have mostly used  device level parameters such as measured photocurrent, voltage, and fill factor to explain the enhanced performance of PSCs with excess \ce{PbI2}. At present, a detailed  understanding of the effect of charge carrier dynamics  and defect formation in the presence of excess \ce{PbI2} is still missing. 

Here, using theory and experiments we perform a comprehensive investigation of the role of surfaces and interfaces present in MAPI thin films when excess \ce{PbI2} is used in the precursor solutions.  Using density functional theory (DFT) calculations, we study how  various surface terminations of MAPI interfaced with \ce{PbI2} lead to different energy level alignments and localization of the energy levels at the valence band maximum (VBM) and the conduction band minimum (CBM) of MAPI. We use our theoretical results to explain the effects of excess \ce{PbI2} and resolve several discrepancies found in the literature.  In particular, we  perform semi-local and hybrid DFT calculations on various possible interfaces between MAPI and \ce{PbI2},  and we use our results to quantify the charge extraction mechanism and formation energy of interfacial defects. In addition, we carry out optical experiments to probe the behavior of charge carriers  at the surfaces of MAPI with various amounts of \ce{PbI2} present in the precursor solution. We find agreement between our theoretical and experimental results, leading  to further clarification of the role of excess \ce{PbI2} in PSCs. Finally, we provide design guidelines for maximizing the beneficial aspects of excess \ce{PbI2} on the performance of PSCs.

\begin{figure}
    \centering
    \includegraphics[scale=0.6]{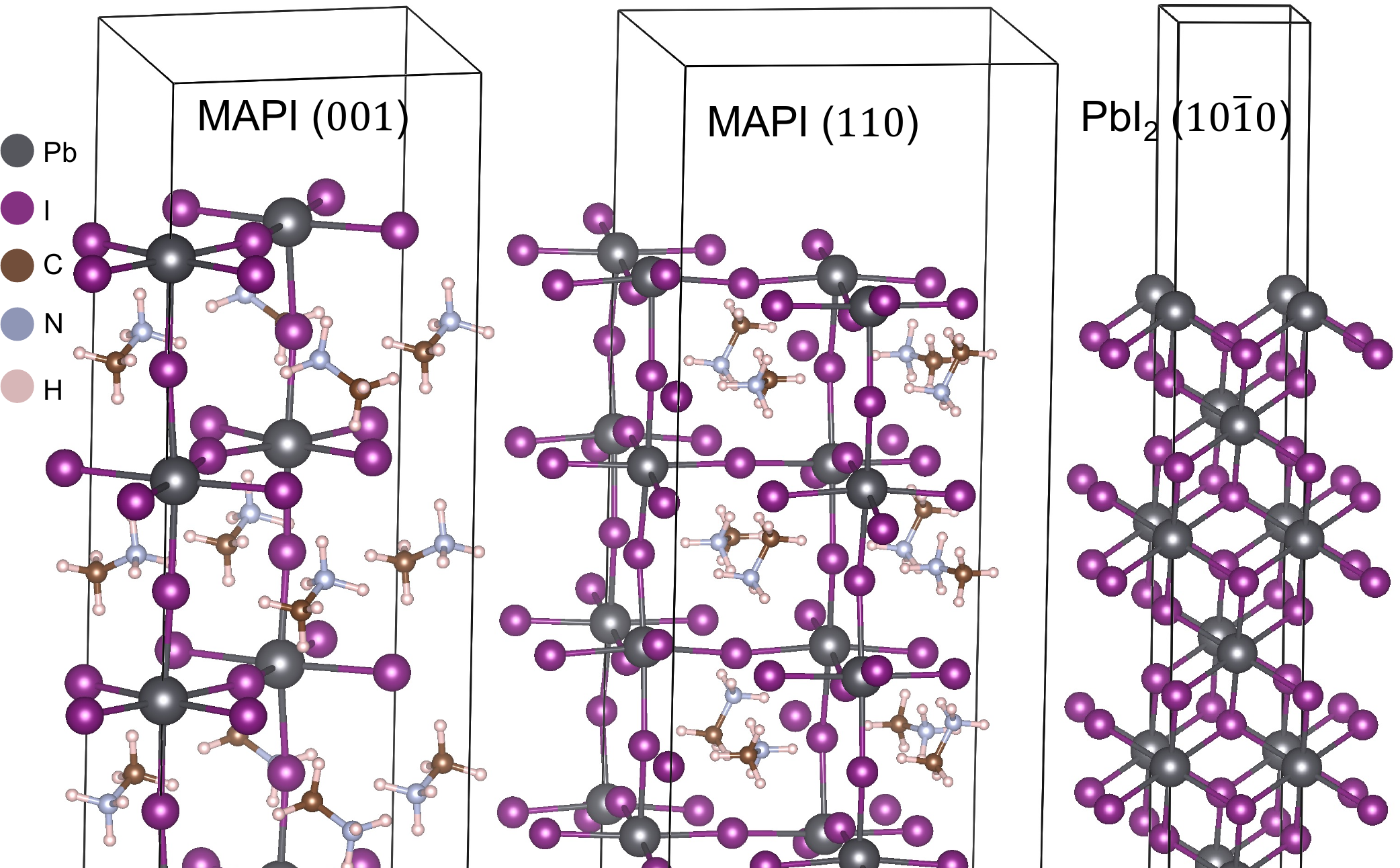}
    \caption{Ball and stick representation of the atomistic structures of the surfaces of MAPI and \ce{PbI2}. The left and middle panels show the dominant \hkl(001) and \hkl(110) surfaces of MAPI respectively, that were used to generate an interface with \hkl(10-10) surface of \ce{PbI2}, shown in the right panel. For MAPI, the layers show \ce{PbI2} and MAI terminations while the \ce{PbI2} surface could be terminated by Pb or I atoms.}
    \label{fig:str}
\end{figure}

\begin{figure}
    \centering
    \includegraphics[scale=0.6]{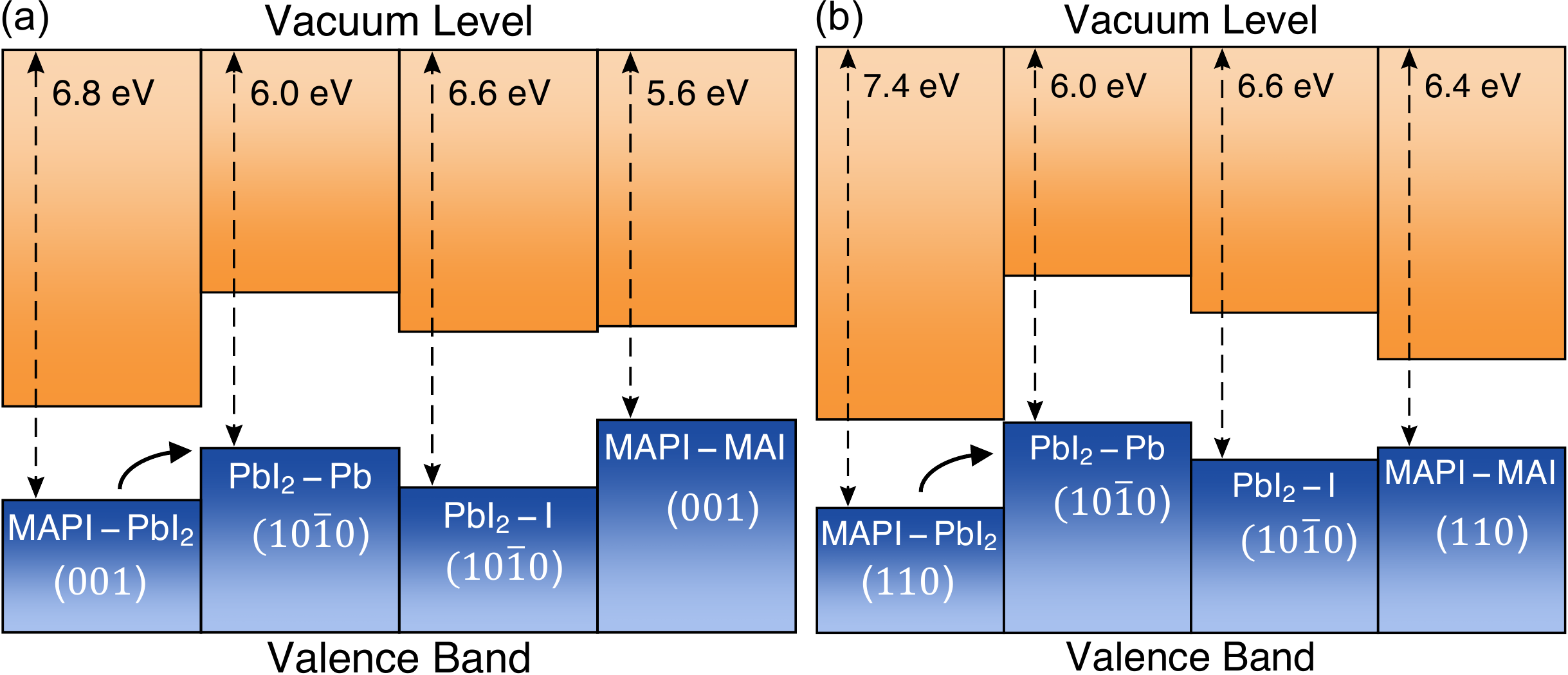}
    \caption{Energy level alignment diagram based on (a) \hkl(001) and (b) \hkl(110) surfaces of MAPI with \ce{PbI2} \hkl(10-10) surface. Both type I and type II alignments are possible depending on the facet and termination of MAPI and \ce{PbI2}. Notably, the \ce{PbI2} termination of MAPI always results in a type II alignment with \ce{PbI2}, while both type I and II alignments are possible with the MAI termination. Generated using band alignment plotting tool, https://github.com/utf/bapt.}
    \label{fig:bandalign}
\end{figure}

We performed hybrid DFT calculations  using the Heyd–Scuseria–Ernzerhof (HSE) functional to obtain accurate energy level alignments between MAPI and \ce{PbI2} as a function of  surface termination. For MAPI, we consider the dominant \hkl(001) and \hkl(110) surfaces~\cite{haruyamaTerminationDependenceTetragonal2014} with \ce{PbI2} and MAI terminations, while for \ce{PbI2}, we consider the \hkl(10-10) surface with Pb and I terminations as shown in Fig.~\ref{fig:str}. The intrinsic differences in the composition and bonding at different surface terminations result in different energy level alignments.
Fig.~\ref{fig:bandalign} shows the energy level alignments between \ce{PbI2} and \hkl(001) and \hkl(110) surfaces of MAPI (a and b respectively). In our notation, the species after the hyphen represents the surface termination, for example, MAPI--\ce{PbI2} refers to the \ce{PbI2} surface termination of MAPI and \ce{PbI2}--I refers to the iodine terminated surface of \ce{PbI2}. First, we focus on MAPI surfaces with \ce{PbI2} termination. We find that  both \hkl(001) and \hkl(110) surfaces have a type II alignment with  \ce{PbI2}, allowing hole transfer from the MAPI valence band to \ce{PbI2} but blocking electron transfer. The energy loss of the hole on moving to the Pb terminated \ce{PbI2} surface is larger than in the case of the I terminated surface. For MAPI surfaces with MAI termination, we obtain a type I interface for MAPI \hkl(001) interfaced  with Pb terminated \ce{PbI2} and type II with I terminated \ce{PbI2}. The type I alignment effectively blocks the transfer of electrons and holes and Pb terminated \ce{PbI2} passivates the surface of MAPI~\cite{chenControllableSelfInducedPassivation2014}. The type II alignment observed here, however, which is due to the CBM of \ce{PbI2} being slightly lower in energy than that of MAPI, allows for the transfer of electrons. For the MAPI \hkl(110) surface, the alignment is of type I with I terminated and type II with Pb terminated \ce{PbI2} surface. The raising of the VBM level of MAI terminated MAPI is consistent with the observations of~\citealt{meggiolaroEnergyLevelTuning2019}. In this case, the VBM level of MAPI is raised above the energy level of the hole transport layer  resulting in blocking hole transfer. 

Our computed energy level diagrams show that both type I and type II alignments may arise,  depending on the terminations of  MAPI and \ce{PbI2} surfaces. As the  amount of \ce{PbI2} is increased and a higher proportion of MAPI surfaces are covered by \ce{PbI2}, a transition to  a primarily type II band alignment is observed which favors hole transfer to \ce{PbI2}. Due to the type II alignment, charge carriers can be more effectively separated, resulting in reduced recombination.  Additionally, in a n-i-p configuration, where the perovskite layer is placed between the electron and hole transport layers, the transfer of holes to the hole transport layer can be enhanced due to the proximity to the \ce{PbI2} terminated surface.  \cite{calloniStabilityOrganicCations2015}.

\begin{figure}
    \centering
    \includegraphics[scale = 0.7]{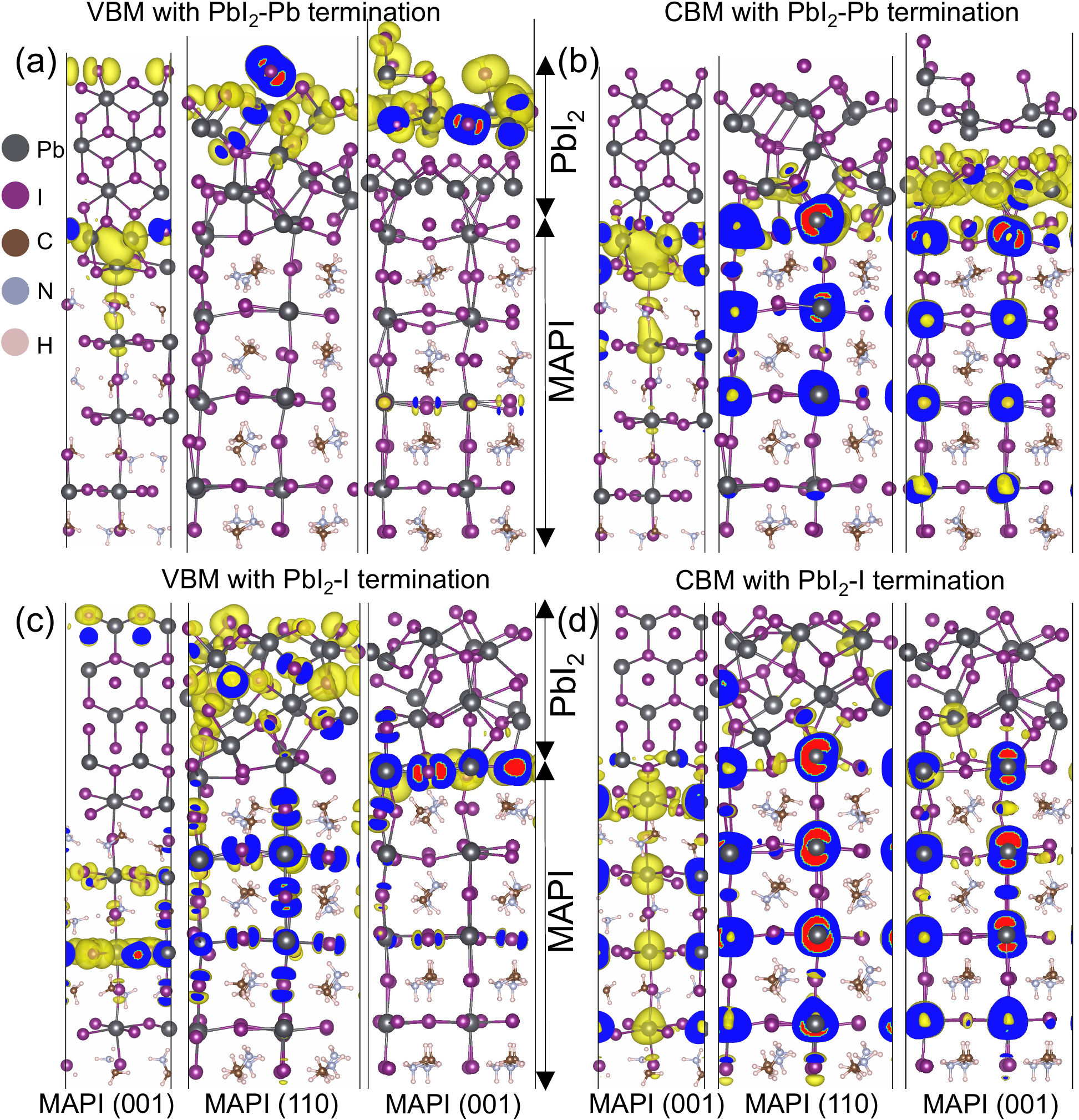}
    \caption{Positions of the valence band maximum (VBM) and conduction band minimum (CBM) for the six interfaces between MAPI  and the \hkl(10-10) surface of \ce{PbI2} considered in this study shown using isosurfaces of electron density for the respective bands.  (a) and (b) show respectively the VBM and CBM positions for interfaces with Pb terminated \ce{PbI2}, while (c) and (d) show the same for interfaces with I terminated \ce{PbI2}. The blue and red colors appear at the locations of the slicing plane. The interfaces are ordered according to Table S1 and generated using three different surfaces of MAPI: \hkl(001), \hkl(110), and \hkl(001) rotated 45 degrees about the axis perpendicular to the interface. The VBM orbitals (a and c) are localized mostly within \ce{PbI2} or at the interface while the CBM orbitals (b and d) lie within the bulk of MAPI. }
    \label{fig:homolumo}
\end{figure}

The energetics of the photogenerated electrons and holes is  not only affected by the surface facet and termination of MAPI and \ce{PbI2}, but also by the  interfacial strain, arising due to lattice mismatch.  
For example, ~\citealt{rothmannAtomicscaleMicrostructureMetal2020a} imaged formamidinium lead iodide (FAPI) solar cells  and showed the existence of coherent interfaces between FAPI and \ce{PbI2} with low lattice distortion. Hence it is interesting to identify  possible commensurate interfaces between MAPI and \ce{PbI2} and study their electronic properties. We characterize  the interface  by the Miller indices of the surfaces of MAPI and \ce{PbI2} in contact with each other, their termination, and individual rotations about the axis perpendicular to the interface. To accommodate the lattice mismatch, we consider several possible multiplicities of the in-plane lattice vectors of MAPI and \ce{PbI2} that result in computationally tractable supercell sizes (Our supercells contain from 120 to 456 atoms, corresponding to 624 and 2272 valence electrons). For all interfaces, we only consider the \ce{PbI2} termination of MAPI in contact with the Pb and I terminations of \ce{PbI2} surfaces to study the effects of excess \ce{PbI2}. We consider six commensurate interfaces between MAPI and \ce{PbI2}, which span the \hkl(001) and \hkl(110) surfaces of MAPI and the \hkl(10-10) surface of \ce{PbI2}. All interfaces are listed in Table S2 in the Supplementary Information. 
While the strain on \ce{PbI2} may be as large as $\sim 10 \%$, we note that along the c axis (perpendicular to the interface) the strain may be easily accommodated  due to the presence of interlayer van der Waals interactions. Further, the energy increase caused by strain can be easily offset by the additional bonding between the surfaces of MAPI and \ce{PbI2} when the \ce{PbI2} film is only a few layers thick~\cite{rooseCriticalAssessmentUse2020}.  The initial distance between the surfaces of MAPI and \ce{PbI2} is set in such a way that Pb-I bonds with bond lengths similar to those found in MAPI ($\sim 3.5 $ {\AA}) are formed connecting the surfaces of MAPI and \ce{PbI2}. Fig.~\ref{fig:homolumo} displays the positions of the VBM and CBM for the six different interfaces between MAPI and \ce{PbI2} considered here. As shown in  Figs.~\ref{fig:homolumo}(a) and (c), we find that the holes are localized within the \ce{PbI2} layers and near the interface, consistent with the band alignment discussed above. This localization is more prominent for the Pb termination. The CBM, as shown in Figs.~\ref{fig:homolumo}(b) and (d), on  the other hand, is localized within the bulk of MAPI. This result further confirms that the effect of excess \ce{PbI2} is to collect the holes, resulting in charge carrier separation and efficient transfer to the hole transport layer. However, if \ce{PbI2} turns out to be completely surrounded by MAPI, it  may be detrimental to the device performance, since its presence will result in the trapping of holes that will have no clear path to the hole transport layer.

Deep traps that limit the power conversion efficiency of PSCs have been shown to concentrate mostly at the surfaces and interfaces present in polycrystalline thin films. ~\cite{niResolvingSpatialEnergetic2020}.
The presence of interfacial defects has been proposed as the cause of the increase of photovoltage at energies below the band gap as the \ce{PbI2} content in the PSC is increased~\cite{supasaiFormationPassivatingCH3NH3PbI32013}. The critical role of surfaces and interfaces is also highlighted in studies on grain size dependence of the perovskite properties~\cite{kimPhotovoltaicPerformancePerovskite2016,xingUltrafastIonMigration2016,renModulatingCrystalGrain2016}. For example, the \ce{I_i^-}/\ce{V_I+} Frenkel pair, where \ce{I_i^-} represents negatively charged interstitial iodine and \ce{V_I+} represents a positively charged iodine vacancy, is under active investigation due to its known role in voltage hysteresis and light-induced photoluminescence enhancement and in voltage decrease in MAPI solar cells~\cite{azpirozDefectsMigrationMethylammonium,eamesIonicTransportHybrid2015,mosconiLightinducedAnnihilationFrenkel2016}. 
~\citealt{meggiolaroFormationSurfaceDefects2019} showed that the formation energy of an \ce{I_i^-}/\ce{V_I+} Frenkel pair decreases drastically from 0.86 eV in the bulk to 0.03 eV at the \ce{PbI2} terminated MAPI surface. This change of energy landscape can result in surface defects dominating the non-radiative recombination characteristics of MAPI PSCs. In MAPI thin films with excess \ce{PbI2}, however, the interfaces between the two materials are expected to play a  more important role in the recombination of charge carriers.  Admittance spectroscopy characterization of perovskite thin films revealed that \ce{PbI2}-rich MAPI PSCs had a lower defect density which resulted in longer carrier lifetimes~\cite{jiangObservationLowerDefect2019}.

Motivated by previous experimental and theoretical studies of defects, we performed DFT calculations of the \ce{I_i^-}/\ce{V_I+} Frenkel pair in MAPI bulk, surfaces, and interfaces between MAPI and \ce{PbI2}. Fig. S1 shows the structure of MAPI containing this defect with the I terminated \ce{PbI2} layer on top. Table~\ref{tab:defect} compares the formation energy of the \ce{I_i^-}/\ce{V_I+} Frenkel pair in MAPI at bulk, surfaces, and interfaces calculated using the Perdew–Burke-Ernzerhof (PBE) functional without spin-orbit coupling (SOC) and the HSE functional with SOC (following common practice in functional comparisons for perovskites, due to cancellation of errors arising from the semi-local functional and spin-orbit effects~\cite{duDensityFunctionalCalculations2015}).
As expected, the defect formation energy  in MAPI decreases from 1.16 eV for the bulk to 0.21 eV for \hkl(001) surface and  is approximately zero  for the \hkl(110) surface.
The defect formation energies for the bulk and \hkl(001) surface agree well with a previous first-principles  investigation~\cite{meggiolaroFormationSurfaceDefects2019}. The defect formation energy for the MAPI \hkl(110) surface obtained with the HSE functional is close to zero (it is in fact slightly negative, -0.01 eV, possibly due to the fact  that the HSE calculation was performed for PBE-optimized geometries),  implying that the \hkl(110) surface might harbor a high density of defects. 
The defect formation energies of MAPI/\ce{PbI2} interfaces follow  two different trends depending on the surface termination of \ce{PbI2} in contact with MAPI. With Pb terminated \ce{PbI2},  the defect formation energy is increased compared to that of the bare MAPI surface. The defect formation energies of the I terminated surfaces, on the other hand, are   comparable to those of the bare MAPI surfaces (both \hkl(001) and \hkl(110) surfaces); hence we conclude that a I-rich  \ce{PbI2} might not provide the beneficial effect of lowering the trap density. Another benefit of using \ce{PbI2} in  I-poor conditions is that the formation of I interstitials in MAPI, that act as deep traps, is inhibited in these conditions. Iodine vacancies, on the other hand, that are formed under I-poor conditions create only shallow traps as shown by hybrid DFT calculations~\cite{duDensityFunctionalCalculations2015,meggiolaroIodineChemistryDetermines2018}.

\begin{table}
  \caption{The formation energy of the \ce{I_i^-}/\ce{V_I+} Frenkel pair in MAPI at bulk, surfaces, and interfaces. The formation energy decreases considerably  at the \ce{PbI2} terminated surfaces, compared to the value in the bulk of MAPI. An interface of MAPI with \ce{PbI2} decreases the defect density under I-poor conditions but increases it under I-rich conditions.}
  \label{tab:defect}
  \begin{tabular}{|ll|l|l|}
    \hline
    \multicolumn{2}{|c}{Structure}  &  \multicolumn{2}{|c|}{Defect Formation Energy} \\
    \hline
      &   & PBE (eV) &  HSE+SOC (eV) \\
      \hline
      Bulk MAPI & &  1.12 & 1.16 \\ 
      MAPI \hkl(001) surface & & 0.03 & 0.21\\
      MAPI \hkl(110) surface & & 0.12 & $\sim0.$\\
      \hline
       \multicolumn{2}{|c|}{Interfaces} & & \\
       MAPI surface & \ce{PbI2} surface & &\\
       \hline
    \hkl(100) & \hkl(10-10)-Pb  & 0.90 & 1.08\\
    \hkl(110) & \hkl(10-10)-Pb  & 0.63 & 0.54\\
    \hkl(100)  & \hkl(10-10)-I &  0.18 &  0.10\\
     \hkl(110) & \hkl(10-10)-I  &   0.04 & 0.03\\
    \hline
  \end{tabular}
\end{table}

We now turn to experiments to validate our  theoretical predictions on energy level alignment, charge extraction, and defect formation energies. The type of band alignment  present at MAPI and PbI2 interfaces influences the surface recombination and charge carrier diffusion at the perovskite interface  and we explore these effects by carrying out optical measurements.

Several experimental studies have attempted to show how excess \ce{PbI2} affects the performance of MAPI PSCs. Few studies, however, probe the effects on charge carrier dynamics at the surface of the film, where the interaction between MAPI and \ce{PbI2} is the most notable. Using a suite of optical characterization methods, we elucidate the behavior of charge carriers at the surface of polycrystalline MAPI thin films with excess amounts of \ce{PbI2} ranging from 0\% (stoichiometric) to 10\% excess, corresponding to typical values considered  in the literature. By combining ultrafast transient reflectance (TR) spectroscopy with time-resolved photoluminescence (TRPL), we determined the diffusion coefficient and surface recombination velocity (SRV) of carriers in MAPI films.

Ultrafast TR spectroscopy, which utilizes an above band gap excitation pump pulse and broadband, white-light probe pulse centered around the band edge, can probe  carrier dynamics at various film depths on a picosecond to nanosecond regime~\cite{Ultrafastprobes,Topandbottom}. By varying the pump energy (i.e. the wavelength of light used in the excitation pulse) different volumes of the sample of interest can be studied due to changes in the absorption coefficient of the material of interest.  At high pump photon energies (and correspondingly shallow pump penetration depths), charge carrier dynamics can primarily be attributed to diffusion of charge carriers away from the surface of the sample, since the probe is only sensitive to within about 10 nms of the surface of the material. However, low- photon pump energies are able to penetrate much farther into the sample film, such that the probed charge carrier behavior is more akin to bulk behavior seen in traditional transient absorption (TA) spectroscopy (Figs. S2 and S3). The recombination dynamics of low photon energy TR (and by extension TA) reflected both the bulk recombination and surface recombination.  Therefore, differences in the TR dynamics taken with different pump excitation wavelengths can be attributed to the SRV since the bulk lifetime remains the same for the material.  By using various pump photon energies, a correlation between the carrier dynamics taken with different excitation energies can be used in a global fit analysis to extract  the SRV as well as the diffusion coefficient. 
Capturing the SRV of carriers in defect tolerant metal halide perovskites using TR, however, can be challenging  due to their low surface recombination rates~\cite{SnaithTRPL}.  The low SRV of carriers in these systems  results in recombination lifetimes larger than the maximum time scale of 2.5 ns analyzed using our TR instrument~\cite{IndividualMobilities}.
Therefore, we use  TRPL to deconvolve the carriers' SRV  from bulk recombination processes by following the recombination lifetime of the carriers as a function of sample thickness for each stoichiometry of interest. The SRV is calculated by plotting the PL lifetime \texttau\textsubscript{PL} as a function of film  thickness L (Fig. S4) and using the relation 1/\texttau\textsubscript{PL} = SRV/L + 1/\texttau\textsubscript{bulk} where  \texttau\textsubscript{bulk} is the bulk lifetime.
Thus, by coupling the two techniques of TR and TRPL, we can extract both the diffusion coefficient  and SRV  of photo-carriers, respectively, for MAPI films with 0, 2.5, 5, and 10\% excess \ce{PbI2}. 
 
\begin{figure}
    \centering
    \includegraphics[scale=1.0]{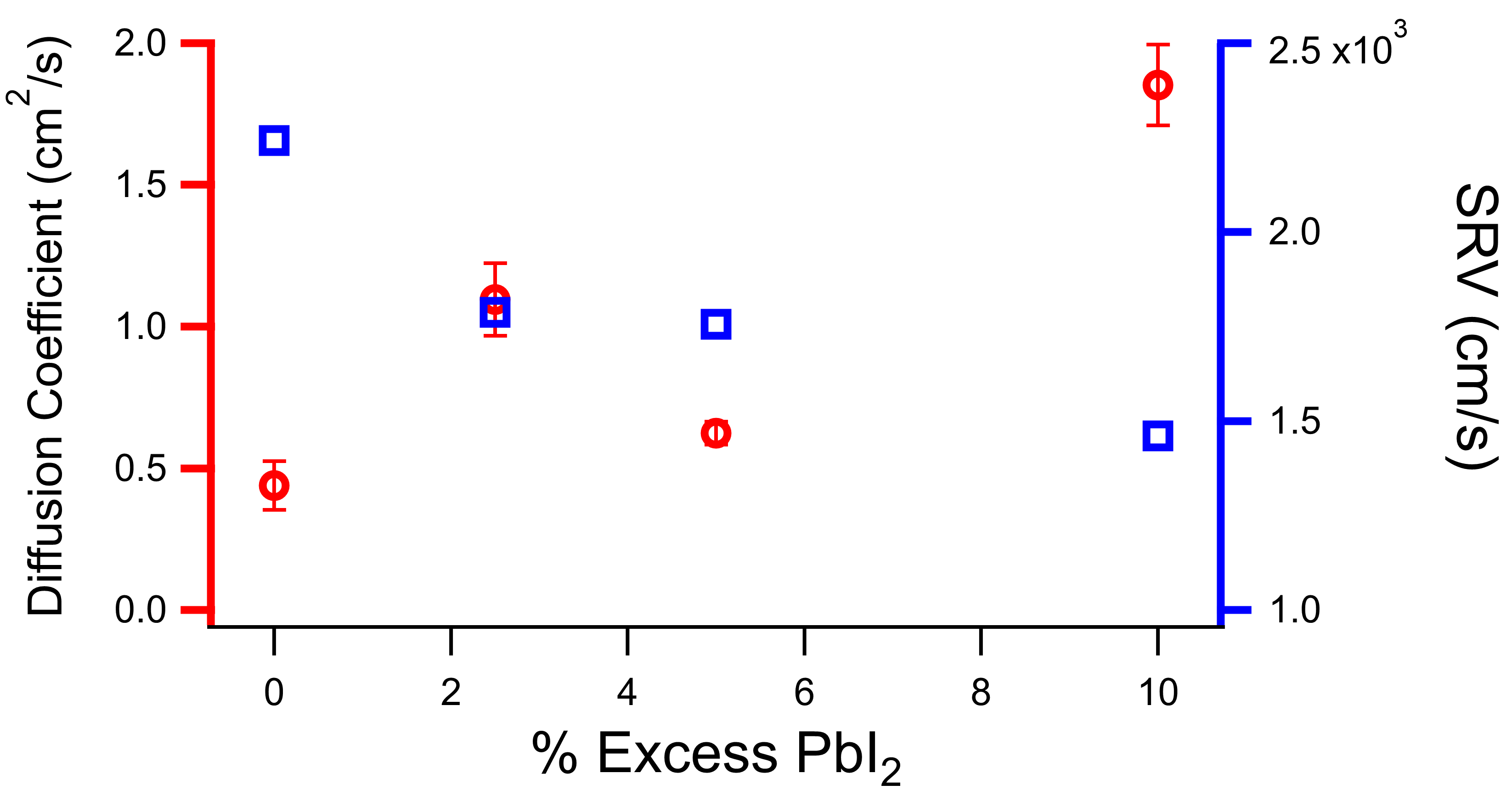}
    \caption{The diffusion coefficient (red, left axis) and the surface recombination velocity (SRV) (blue, right axis) as a function of excess \ce{PbI2} in the MAPI films.  The diffusion coefficient is obtained from the global fit analysis performed on the transient reflectance data (Fig. S3).  Increasing amounts of \ce{PbI2} yields increasing diffusion coefficients, indicating that fewer bulk defects are present.  However, in the case of 5\% excess \ce{PbI2}, the diffusion coefficient is similar to the stoichiometric instance.  The SRV decreases with increasing amounts of \ce{PbI2}, indicating surface passivation.}
    \label{fig:diffusion}
\end{figure}
The diffusion coefficient and SRV are shown in Fig.~\ref{fig:diffusion}. The diffusion coefficient increases with increasing amounts of \ce{PbI2}, with 10\% excess showing the highest diffusion coefficient of photogenerated charge carriers. At 5\% excess \ce{PbI2}, there is a drop in the diffusion coefficient with a value close to that for the stoichiometric sample. 
The overall trend observed for the diffusion coefficient and SRV indicates that increasing the amount of \ce{PbI2} leads to better charge carrier  and lower SRV which in turn should translate to an enhanced solar cell performance. Indeed, multiple studies show that excess \ce{PbI2} leads to better device performance, with a peak performance around 10\% excess \ce{PbI2}~\cite{merdasaImpactExcessLead2019,rooseCriticalAssessmentUse2020,UnravelingShi}. It was previously suggested that the presence of excess \ce{PbI2} boosted device performance purely by passivating surface trap states created by halide ion vacancies~\cite{jacobssonUnreactedPbI2DoubleEdged2016}. Since the diffusion coefficient is a measure of the ability of charge carriers to diffuse away from the surface of the film towards the bulk, this trend indicates that excess \ce{PbI2} does not just influence the surface of MAPI films. Our results show that excess \ce{PbI2} increases charge carrier diffusion away from the surface, as well. This may be due to faster transport (fewer defects) at grain boundaries and/or a lower number of grain boundaries in the excess \ce{PbI2} samples, as has been proposed~\cite{jacobssonUnreactedPbI2DoubleEdged2016}. However, it is important to note that the samples with different stoichiometries investigated here have similar grain sizes   (Fig. S5); therefore any change in charge carrier behavior can be attributed to other phenomena besides carrier recombination at grain boundaries.  The reduced defect density at Pb terminated \ce{PbI2} interface with MAPI predicted by theory and shown in Fig. \ref{fig:homolumo}, may additionally contribute to higher diffusion of charge carriers.  Additionally, for all stoichiometries studied, the \texttau\textsubscript{bulk} of the materials are the same, indicating that the changes to the PL lifetime of the samples is due solely to change in the surface recombination.  This observation provides further support that excess \ce{PbI2} affects the surface of the perovskite, and therefore the interface between the perovskite and \ce{PbI2}.

The SRV is the propensity of charge carriers to recombine at the surface, with larger values indicating a large probability of recombination due to a higher density of recombination centers~\cite{Topandbottom}. 
At 0\% excess \ce{PbI2}, the SRV is $2.24\times 10^3$ m/s and it  drops across the other stoichiometries down to $1.46 \times 10^3$ cm/s for the 10\% excess sample.  This clear trend indicates that films with excess \ce{PbI2} have fewer defects and recombination centers at the surface than films with stoichiometric amounts of \ce{PbI2}. The  lower SRV can be due to the interplay of a combination of factors.  One source of the drop in SRV can be  the passivation of dangling bonds at surfaces of MAPI and changes in the defect densities at surfaces and interfaces~\cite{Topandbottom}.  Additionally, excess \ce{PbI2} can act as a hole sink, creating a barrier for holes to reach the surface of the perovskite. This energy barrier is quantified by the difference in energy levels of the VBM between Pb terminated \ce{PbI2} and MAPI shown in Fig.~\ref{fig:bandalign}.  When the holes quickly diffuse to excess \ce{PbI2} but the electrons remain within the perovskite, reduced recombination and larger charge carrier lifetimes are expected, which is reflected in the change in SRV.
It is interesting to note that while the diffusion coefficient of the 5\% excess \ce{PbI2} sample is close to that of the stoichiometric sample, the SRV is much lower. Again, while we cannot distinguish a single cause of the drop in diffusion coefficient from 2.5\% to 5\% excess \ce{PbI2}, we can identify several important factors leading to such drop.  First, the holes from MAPI may get trapped and immobilized within \ce{PbI2}; hence they do not diffuse throughout the perovskite to recombine with electrons. Such a situation may occur, for example,  when pockets of \ce{PbI2} are completely surrounded by MAPI~\cite{rooseCriticalAssessmentUse2020}.
The drop in the diffusion coefficient may also stem from the increased concentration of Frenkel pairs in MAPI due to the  preferred formation of  MAPI interfaces  with I terminated \ce{PbI2}, when   the excess amount is increased from 2.5\% to 5\%.  Our theoretical results presented earlier indicate that for I terminated surfaces, Frenkel defects can form more easily than with other terminations.
The \ce{I_i^-} interstitial from these Frenkel defects generated at the interface can easily migrate to the bulk of MAPI due to the low migration barrier of $\sim$0.1 eV~\cite{meggiolaroFormationSurfaceDefects2019} and 
act as an effective hole trap within the bulk~\cite{zhangIodineInterstitialsCause2020}, thereby reducing the diffusion coefficient while not  affecting the SRV.
However, between 5\% and 10\%  excess, we posit that Pb terminations become more frequent than I terminations, allowing for charge extraction into the \ce{PbI2} layer. This can be seen through the drop in SRV since charge transfer becomes more prominent than recombination. This shift towards more charge transfer than recombination is also seen in the band alignment at the interface, as predicted by theory and shown in Fig. \ref{fig:bandalign}.

In summary, by combining optical spectroscopy measurements  with  first-principles electronic structure calculations  of the  interfaces between  MAPI and \ce{PbI2}, we  provided microscopic  insights into the role of excess \ce{PbI2} in PSCs. Consistent with previous investigations, our results point at the key role  of surface terminations and interfaces in  determining  the properties of hybrid perovskites~\cite{huangObservationSpatiallyResolved2021,myungRashbaDresselhausEffect2018}.  We found  that excess \ce{PbI2} on MAPI does not just provide passivation of the MAPI surface. Depending on whether the metal or the halide terminate the \ce{PbI2} surface, different band alignments with the MAPI are possible and, importantly,  different concentrations of Frenkel defects  may be present, which in turn affect the electronic structure of the interface. Our findings  illustrate the important interplay between Frenkel defect formation and the microscopic structure of excess \ce{PbI2}, especially its surface terminations, in determining the enhancement or decrease of the photoluminescence intensity of photo-excited MAPI~\cite{mottiControllingCompetingPhotochemical2019}.
We conclude that while excess \ce{PbI2} provides a facile route to passivating MAPI, further enhancement of device performance would be better realized through the addition of other Pb-containing additives.  For example, the addition of \ce{PbSO4}~\cite{yangStabilizingHalidePerovskite2019} could passivate the MAPI surface   with Pb terminated surfaces of the excess layer without the detrimental effects that additional \ce{I2} would cause. 
While we recognize that perovskite thin films are extremely sensitive to  processing conditions~\cite{HerzChargeCarrierPerspective}, our work points at an important, general design guideline to improve device efficiency. We show that the effect of \ce{PbI2} on device performance is not monomodal,   and that \ce{PbI2} is not an ideal additive for MAPI based devices. We recommend that other additives be explored to enhance PSC performance, specifically, ones that allow for the benefits of Pb terminations such as reduced trap densities without the  detrimental effects of excess \ce{I2}.

\begin{acknowledgement}

We thank Dr. Volker Blum for helpful discussions. The study is supported as part of the Center for Hybrid Organic Inorganic Semiconductor for Energy (CHOISE), an Energy Frontier Research Center funded by the Office of Basic Energy Sciences, Office of Science within the U.S. Department of Energy. This work was authored in part by the Alliance for Sustainable Energy, LLC, the manager and operator of the National Renewable Energy Laboratory for DOE under contract number DE-AC36-08GO28308. The views expressed in the Article do not necessarily represent the views of the DOE or the U.S. Government. The U.S. Government retains and the publisher, by accepting the article for publication, acknowledges that the U.S. Government retains a nonexclusive, paid-up, irrevocable, worldwide license to publish or reproduce the published form of this work, or allow others to do so, for U.S. Government purposes.  We acknowledge the National Energy Research Scientific Computing Center (NERSC), a U.S. Department of Energy Office of Science User Facility located at Lawrence Berkeley National Laboratory, operated under Contract No. DE-AC02-05CH11231 and the Bridges-2 system supported by NSF grant number ACI-1445606 which is a part of Extreme Science and Engineering Discovery Environment (XSEDE), funded by NSF grant number ACI-1548562 for computational resources.

\end{acknowledgement}

\begin{suppinfo}

Details of computational and experimental methods and a compilation of all interfaces considered in this work.

\end{suppinfo}

\bibliography{allrefs}

\providecommand{\latin}[1]{#1}
\makeatletter
\providecommand{\doi}
  {\begingroup\let\do\@makeother\dospecials
  \catcode`\{=1 \catcode`\}=2 \doi@aux}
\providecommand{\doi@aux}[1]{\endgroup\texttt{#1}}
\makeatother
\providecommand*\mcitethebibliography{\thebibliography}
\csname @ifundefined\endcsname{endmcitethebibliography}
  {\let\endmcitethebibliography\endthebibliography}{}
\begin{mcitethebibliography}{43}
\providecommand*\natexlab[1]{#1}
\providecommand*\mciteSetBstSublistMode[1]{}
\providecommand*\mciteSetBstMaxWidthForm[2]{}
\providecommand*\mciteBstWouldAddEndPuncttrue
  {\def\EndOfBibitem{\unskip.}}
\providecommand*\mciteBstWouldAddEndPunctfalse
  {\let\EndOfBibitem\relax}
\providecommand*\mciteSetBstMidEndSepPunct[3]{}
\providecommand*\mciteSetBstSublistLabelBeginEnd[3]{}
\providecommand*\EndOfBibitem{}
\mciteSetBstSublistMode{f}
\mciteSetBstMaxWidthForm{subitem}{(\alph{mcitesubitemcount})}
\mciteSetBstSublistLabelBeginEnd
  {\mcitemaxwidthsubitemform\space}
  {\relax}
  {\relax}

\bibitem[Green \latin{et~al.}(2014)Green, Ho-Baillie, and
  Snaith]{Green2014emergence}
Green,~M.~A.; Ho-Baillie,~A.; Snaith,~H.~J. The emergence of perovskite solar
  cells. \emph{Nat. Photonics} \textbf{2014}, \emph{8}, 506--514\relax
\mciteBstWouldAddEndPuncttrue
\mciteSetBstMidEndSepPunct{\mcitedefaultmidpunct}
{\mcitedefaultendpunct}{\mcitedefaultseppunct}\relax
\EndOfBibitem
\bibitem[Huang \latin{et~al.}(2017)Huang, Yuan, Shao, and
  Yan]{huangUnderstandingPhysicalProperties2017}
Huang,~J.; Yuan,~Y.; Shao,~Y.; Yan,~Y. Understanding the Physical Properties of
  Hybrid Perovskites for Photovoltaic Applications. \emph{Nat. Rev. Mater.}
  \textbf{2017}, \emph{2}, 17042\relax
\mciteBstWouldAddEndPuncttrue
\mciteSetBstMidEndSepPunct{\mcitedefaultmidpunct}
{\mcitedefaultendpunct}{\mcitedefaultseppunct}\relax
\EndOfBibitem
\bibitem[Yoo \latin{et~al.}(2021)Yoo, Seo, Chua, Park, Lu, Rotermund, Kim,
  Moon, Jeon, {Correa-Baena}, Bulovi{\'c}, Shin, Bawendi, and
  Seo]{yooEfficientPerovskiteSolar2021}
Yoo,~J.~J.; Seo,~G.; Chua,~M.~R.; Park,~T.~G.; Lu,~Y.; Rotermund,~F.;
  Kim,~Y.-K.; Moon,~C.~S.; Jeon,~N.~J.; {Correa-Baena},~J.-P.; Bulovi{\'c},~V.;
  Shin,~S.~S.; Bawendi,~M.~G.; Seo,~J. Efficient Perovskite Solar Cells via
  Improved Carrier Management. \emph{Nature} \textbf{2021}, \emph{590},
  587--593\relax
\mciteBstWouldAddEndPuncttrue
\mciteSetBstMidEndSepPunct{\mcitedefaultmidpunct}
{\mcitedefaultendpunct}{\mcitedefaultseppunct}\relax
\EndOfBibitem
\bibitem[Li \latin{et~al.}(2017)Li, Wang, Deschler, Gao, Friend, and
  Cheetham]{liChemicallyDiverseMultifunctional2017}
Li,~W.; Wang,~Z.; Deschler,~F.; Gao,~S.; Friend,~R.~H.; Cheetham,~A.~K.
  Chemically Diverse and Multifunctional Hybrid Organic\textendash Inorganic
  Perovskites. \emph{Nat. Rev. Mater.} \textbf{2017}, \emph{2}, 16099\relax
\mciteBstWouldAddEndPuncttrue
\mciteSetBstMidEndSepPunct{\mcitedefaultmidpunct}
{\mcitedefaultendpunct}{\mcitedefaultseppunct}\relax
\EndOfBibitem
\bibitem[{National Renewable Energy Laboratory, Golden,
  Colorado}(2021)]{NRELChart}
{National Renewable Energy Laboratory, Golden, Colorado}, Best Research-Cell
  Efficiency Chart. \emph{NREL} \textbf{2021}, \relax
\mciteBstWouldAddEndPunctfalse
\mciteSetBstMidEndSepPunct{\mcitedefaultmidpunct}
{}{\mcitedefaultseppunct}\relax
\EndOfBibitem
\bibitem[Jacobsson \latin{et~al.}(2016)Jacobsson, {Correa-Baena}, Anaraki,
  Philippe, Stranks, Bouduban, Tress, Schenk, and
  Moser]{jacobssonUnreactedPbI2DoubleEdged2016}
Jacobsson,~T.~J.; {Correa-Baena},~J.-P.; Anaraki,~E.~H.; Philippe,~B.;
  Stranks,~S.~D.; Bouduban,~M. E.~F.; Tress,~W.; Schenk,~K.; Moser,~J.-E.
  Unreacted {{PbI2}} as a {{Double-Edged Sword}} for {{Enhancing}} the
  {{Performance}} of {{Perovskite Solar Cells}}. \emph{J. Am. Chem. Soc.}
  \textbf{2016}, 13\relax
\mciteBstWouldAddEndPuncttrue
\mciteSetBstMidEndSepPunct{\mcitedefaultmidpunct}
{\mcitedefaultendpunct}{\mcitedefaultseppunct}\relax
\EndOfBibitem
\bibitem[Chen \latin{et~al.}(2014)Chen, Zhou, Song, Luo, Hong, Duan, Dou, Liu,
  and Yang]{chenControllableSelfInducedPassivation2014}
Chen,~Q.; Zhou,~H.; Song,~T.-B.; Luo,~S.; Hong,~Z.; Duan,~H.-S.; Dou,~L.;
  Liu,~Y.; Yang,~Y. Controllable {{Self-Induced Passivation}} of {{Hybrid Lead
  Iodide Perovskites}} toward {{High Performance Solar Cells}}. \emph{Nano
  Lett.} \textbf{2014}, \emph{14}, 4158--4163\relax
\mciteBstWouldAddEndPuncttrue
\mciteSetBstMidEndSepPunct{\mcitedefaultmidpunct}
{\mcitedefaultendpunct}{\mcitedefaultseppunct}\relax
\EndOfBibitem
\bibitem[Supasai \latin{et~al.}(2013)Supasai, Rujisamphan, Ullrich,
  Chemseddine, and Dittrich]{supasaiFormationPassivatingCH3NH3PbI32013}
Supasai,~T.; Rujisamphan,~N.; Ullrich,~K.; Chemseddine,~A.; Dittrich,~T.
  Formation of a Passivating {{CH3NH3PbI3}}/{{PbI2}} Interface during Moderate
  Heating of {{CH3NH3PbI3}} Layers. \emph{Appl. Phys. Lett.} \textbf{2013},
  \emph{103}, 183906\relax
\mciteBstWouldAddEndPuncttrue
\mciteSetBstMidEndSepPunct{\mcitedefaultmidpunct}
{\mcitedefaultendpunct}{\mcitedefaultseppunct}\relax
\EndOfBibitem
\bibitem[Roose \latin{et~al.}(2020)Roose, Dey, Chiang, Friend, and
  Stranks]{rooseCriticalAssessmentUse2020}
Roose,~B.; Dey,~K.; Chiang,~Y.-H.; Friend,~R.~H.; Stranks,~S.~D. Critical
  {{Assessment}} of the {{Use}} of {{Excess Lead Iodide}} in {{Lead Halide
  Perovskite Solar Cells}}. \emph{J. Phys. Chem. Lett.} \textbf{2020},
  \emph{11}, 6505--6512\relax
\mciteBstWouldAddEndPuncttrue
\mciteSetBstMidEndSepPunct{\mcitedefaultmidpunct}
{\mcitedefaultendpunct}{\mcitedefaultseppunct}\relax
\EndOfBibitem
\bibitem[{Rold{\'a}n-Carmona} \latin{et~al.}(2015){Rold{\'a}n-Carmona}, Gratia,
  Zimmermann, Grancini, Gao, Graetzel, and
  Nazeeruddin]{roldan-carmonaHighEfficiencyMethylammonium2015}
{Rold{\'a}n-Carmona},~C.; Gratia,~P.; Zimmermann,~I.; Grancini,~G.; Gao,~P.;
  Graetzel,~M.; Nazeeruddin,~M.~K. High Efficiency Methylammonium Lead
  Triiodide Perovskite Solar Cells: The Relevance of Non-Stoichiometric
  Precursors. \emph{Energy Environ. Sci.} \textbf{2015}, \emph{8},
  3550--3556\relax
\mciteBstWouldAddEndPuncttrue
\mciteSetBstMidEndSepPunct{\mcitedefaultmidpunct}
{\mcitedefaultendpunct}{\mcitedefaultseppunct}\relax
\EndOfBibitem
\bibitem[Barb{\'e} \latin{et~al.}(2018)Barb{\'e}, Newman, Lilliu, Kumar, Lee,
  Charbonneau, Rodenburg, Lidzey, and Tsoi]{barbeLocalizedEffectPbI22018}
Barb{\'e},~J.; Newman,~M.; Lilliu,~S.; Kumar,~V.; Lee,~H. K.~H.;
  Charbonneau,~C.; Rodenburg,~C.; Lidzey,~D.; Tsoi,~W.~C. Localized Effect of
  {{PbI2}} Excess in Perovskite Solar Cells Probed by High-Resolution
  Chemical\textendash Optoelectronic Mapping. \emph{J. Mater. Chem. A}
  \textbf{2018}, \emph{6}, 23010--23018\relax
\mciteBstWouldAddEndPuncttrue
\mciteSetBstMidEndSepPunct{\mcitedefaultmidpunct}
{\mcitedefaultendpunct}{\mcitedefaultseppunct}\relax
\EndOfBibitem
\bibitem[Merdasa \latin{et~al.}(2019)Merdasa, Kiligaridis, Rehermann,
  {Abdi-Jalebi}, St{\"o}ber, Louis, Gerhard, Stranks, Unger, and
  Scheblykin]{merdasaImpactExcessLead2019}
Merdasa,~A.; Kiligaridis,~A.; Rehermann,~C.; {Abdi-Jalebi},~M.; St{\"o}ber,~J.;
  Louis,~B.; Gerhard,~M.; Stranks,~S.~D.; Unger,~E.~L.; Scheblykin,~I.~G.
  Impact of {{Excess Lead Iodide}} on the {{Recombination Kinetics}} in {{Metal
  Halide Perovskites}}. \emph{ACS Energy Lett.} \textbf{2019}, \emph{4},
  1370--1378\relax
\mciteBstWouldAddEndPuncttrue
\mciteSetBstMidEndSepPunct{\mcitedefaultmidpunct}
{\mcitedefaultendpunct}{\mcitedefaultseppunct}\relax
\EndOfBibitem
\bibitem[Jiang \latin{et~al.}(2019)Jiang, Wu, Zhou, and
  Wang]{jiangObservationLowerDefect2019}
Jiang,~M.; Wu,~Y.; Zhou,~Y.; Wang,~Z. Observation of Lower Defect Density
  Brought by Excess {{PbI2}} in {{CH3NH3PbI3}} Solar Cells. \emph{AIP Adv.}
  \textbf{2019}, \emph{9}, 085301\relax
\mciteBstWouldAddEndPuncttrue
\mciteSetBstMidEndSepPunct{\mcitedefaultmidpunct}
{\mcitedefaultendpunct}{\mcitedefaultseppunct}\relax
\EndOfBibitem
\bibitem[Calloni \latin{et~al.}(2015)Calloni, Abate, Bussetti, Berti,
  Yivlialin, Ciccacci, and Du{\`o}]{calloniStabilityOrganicCations2015}
Calloni,~A.; Abate,~A.; Bussetti,~G.; Berti,~G.; Yivlialin,~R.; Ciccacci,~F.;
  Du{\`o},~L. Stability of {{Organic Cations}} in {{Solution-Processed CH}}
  {\textsubscript{3}} {{NH}} {\textsubscript{3}} {{PbI}} {\textsubscript{3}}
  {{Perovskites}}: {{Formation}} of {{Modified Surface Layers}}. \emph{J. Phys.
  Chem. C} \textbf{2015}, \emph{119}, 21329--21335\relax
\mciteBstWouldAddEndPuncttrue
\mciteSetBstMidEndSepPunct{\mcitedefaultmidpunct}
{\mcitedefaultendpunct}{\mcitedefaultseppunct}\relax
\EndOfBibitem
\bibitem[Gujar \latin{et~al.}(2018)Gujar, Unger, Sch{\"o}nleber, Fried, Panzer,
  {van Smaalen}, K{\"o}hler, and Thelakkat]{gujarRolePbICH2018}
Gujar,~T.~P.; Unger,~T.; Sch{\"o}nleber,~A.; Fried,~M.; Panzer,~F.; {van
  Smaalen},~S.; K{\"o}hler,~A.; Thelakkat,~M. The Role of {{PbI}}
  {\textsubscript{2}} in {{CH}} {\textsubscript{3}} {{NH}} {\textsubscript{3}}
  {{PbI}} {\textsubscript{3}} Perovskite Stability, Solar Cell Parameters and
  Device Degradation. \emph{Phys. Chem. Chem. Phys.} \textbf{2018}, \emph{20},
  605--614\relax
\mciteBstWouldAddEndPuncttrue
\mciteSetBstMidEndSepPunct{\mcitedefaultmidpunct}
{\mcitedefaultendpunct}{\mcitedefaultseppunct}\relax
\EndOfBibitem
\bibitem[Wang \latin{et~al.}(2017)Wang, Hao, Han, Yu, Qin, Zhang, Guo, Ai, and
  Zhang]{wangAdverseEffectsExcess2017}
Wang,~H.-Y.; Hao,~M.-Y.; Han,~J.; Yu,~M.; Qin,~Y.; Zhang,~P.; Guo,~Z.-X.;
  Ai,~X.-C.; Zhang,~J.-P. Adverse {{Effects}} of {{Excess Residual PbI2}} on
  {{Photovoltaic Performance}}, {{Charge Separation}}, and {{Trap-State
  Properties}} in {{Mesoporous Structured Perovskite Solar Cells}}. \emph{Chem.
  - Eur. J.} \textbf{2017}, \emph{23}, 3986--3992\relax
\mciteBstWouldAddEndPuncttrue
\mciteSetBstMidEndSepPunct{\mcitedefaultmidpunct}
{\mcitedefaultendpunct}{\mcitedefaultseppunct}\relax
\EndOfBibitem
\bibitem[Liu \latin{et~al.}(2016)Liu, Dong, Wong, Djuri{\v s}i{\'c}, Ng, Ren,
  Shen, Surya, Chan, Wang, Ng, Liao, Li, Shih, Wei, Su, and
  Dai]{liuExcessPbI2Beneficial2016}
Liu,~F. \latin{et~al.}  Is {{Excess PbI2 Beneficial}} for {{Perovskite Solar
  Cell Performance}}? \emph{Adv. Energy Mater.} \textbf{2016}, \emph{6},
  1502206\relax
\mciteBstWouldAddEndPuncttrue
\mciteSetBstMidEndSepPunct{\mcitedefaultmidpunct}
{\mcitedefaultendpunct}{\mcitedefaultseppunct}\relax
\EndOfBibitem
\bibitem[J.~Dahlman \latin{et~al.}(2021)J.~Dahlman, J.~Kubicki, and
  Manjunatha~Reddy]{j.dahlmanInterfacesMetalHalide2021}
J.~Dahlman,~C.; J.~Kubicki,~D.; Manjunatha~Reddy,~G.~N. Interfaces in Metal
  Halide Perovskites Probed by Solid-State {{NMR}} Spectroscopy. \emph{J.
  Mater. Chem.A} \textbf{2021}, \relax
\mciteBstWouldAddEndPunctfalse
\mciteSetBstMidEndSepPunct{\mcitedefaultmidpunct}
{}{\mcitedefaultseppunct}\relax
\EndOfBibitem
\bibitem[Ni \latin{et~al.}(2020)Ni, Bao, Liu, Jiang, Wu, Chen, Dai, Chen,
  Hartweg, Yu, Holman, and Huang]{niResolvingSpatialEnergetic2020}
Ni,~Z.; Bao,~C.; Liu,~Y.; Jiang,~Q.; Wu,~W.-Q.; Chen,~S.; Dai,~X.; Chen,~B.;
  Hartweg,~B.; Yu,~Z.; Holman,~Z.; Huang,~J. Resolving Spatial and Energetic
  Distributions of Trap States in Metal Halide Perovskite Solar Cells.
  \emph{Science} \textbf{2020}, \emph{367}, 1352--1358\relax
\mciteBstWouldAddEndPuncttrue
\mciteSetBstMidEndSepPunct{\mcitedefaultmidpunct}
{\mcitedefaultendpunct}{\mcitedefaultseppunct}\relax
\EndOfBibitem
\bibitem[Haruyama \latin{et~al.}(2014)Haruyama, Sodeyama, Han, and
  Tateyama]{haruyamaTerminationDependenceTetragonal2014}
Haruyama,~J.; Sodeyama,~K.; Han,~L.; Tateyama,~Y. Termination {{Dependence}} of
  {{Tetragonal CH}} {\textsubscript{3}} {{NH}} {\textsubscript{3}} {{PbI}}
  {\textsubscript{3}} {{Surfaces}} for {{Perovskite Solar Cells}}. \emph{J.
  Phys. Chem. Lett.} \textbf{2014}, \emph{5}, 2903--2909\relax
\mciteBstWouldAddEndPuncttrue
\mciteSetBstMidEndSepPunct{\mcitedefaultmidpunct}
{\mcitedefaultendpunct}{\mcitedefaultseppunct}\relax
\EndOfBibitem
\bibitem[Meggiolaro \latin{et~al.}(2019)Meggiolaro, Mosconi, Proppe,
  {Quintero-Bermudez}, Kelley, Sargent, and
  De~Angelis]{meggiolaroEnergyLevelTuning2019}
Meggiolaro,~D.; Mosconi,~E.; Proppe,~A.~H.; {Quintero-Bermudez},~R.;
  Kelley,~S.~O.; Sargent,~E.~H.; De~Angelis,~F. Energy {{Level Tuning}} at the
  {{MAPbI}} {\textsubscript{3}} {{Perovskite}}/{{Contact Interface Using
  Chemical Treatment}}. \emph{ACS Energy Lett.} \textbf{2019}, \emph{4},
  2181--2184\relax
\mciteBstWouldAddEndPuncttrue
\mciteSetBstMidEndSepPunct{\mcitedefaultmidpunct}
{\mcitedefaultendpunct}{\mcitedefaultseppunct}\relax
\EndOfBibitem
\bibitem[Rothmann \latin{et~al.}(2020)Rothmann, Kim, Borchert, Lohmann,
  O'Leary, Sheader, Clark, Snaith, Johnston, Nellist, and
  Herz]{rothmannAtomicscaleMicrostructureMetal2020a}
Rothmann,~M.~U.; Kim,~J.~S.; Borchert,~J.; Lohmann,~K.~B.; O'Leary,~C.~M.;
  Sheader,~A.~A.; Clark,~L.; Snaith,~H.~J.; Johnston,~M.~B.; Nellist,~P.~D.;
  Herz,~L.~M. Atomic-Scale Microstructure of Metal Halide Perovskite.
  \emph{Science} \textbf{2020}, \emph{370}\relax
\mciteBstWouldAddEndPuncttrue
\mciteSetBstMidEndSepPunct{\mcitedefaultmidpunct}
{\mcitedefaultendpunct}{\mcitedefaultseppunct}\relax
\EndOfBibitem
\bibitem[Kim \latin{et~al.}(2016)Kim, Ohkita, Benten, and
  Ito]{kimPhotovoltaicPerformancePerovskite2016}
Kim,~H.~D.; Ohkita,~H.; Benten,~H.; Ito,~S. Photovoltaic {{Performance}} of
  {{Perovskite Solar Cells}} with {{Different Grain Sizes}}. \emph{Adv. Mater.}
  \textbf{2016}, \emph{28}, 917--922\relax
\mciteBstWouldAddEndPuncttrue
\mciteSetBstMidEndSepPunct{\mcitedefaultmidpunct}
{\mcitedefaultendpunct}{\mcitedefaultseppunct}\relax
\EndOfBibitem
\bibitem[Xing \latin{et~al.}(2016)Xing, Wang, Dong, Yuan, Fang, and
  Huang]{xingUltrafastIonMigration2016}
Xing,~J.; Wang,~Q.; Dong,~Q.; Yuan,~Y.; Fang,~Y.; Huang,~J. Ultrafast Ion
  Migration in Hybrid Perovskite Polycrystalline Thin Films under Light and
  Suppression in Single Crystals. \emph{Phys. Chem. Chem. Phys.} \textbf{2016},
  \emph{18}, 30484--30490\relax
\mciteBstWouldAddEndPuncttrue
\mciteSetBstMidEndSepPunct{\mcitedefaultmidpunct}
{\mcitedefaultendpunct}{\mcitedefaultseppunct}\relax
\EndOfBibitem
\bibitem[Ren \latin{et~al.}(2016)Ren, Yang, Yang, Zhang, Cui, Liu, Wei, Fan,
  and Liu]{renModulatingCrystalGrain2016}
Ren,~X.; Yang,~Z.; Yang,~D.; Zhang,~X.; Cui,~D.; Liu,~Y.; Wei,~Q.; Fan,~H.;
  Liu,~S.~F. Modulating Crystal Grain Size and Optoelectronic Properties of
  Perovskite Films for Solar Cells by Reaction Temperature. \emph{Nanoscale}
  \textbf{2016}, \emph{8}, 3816--3822\relax
\mciteBstWouldAddEndPuncttrue
\mciteSetBstMidEndSepPunct{\mcitedefaultmidpunct}
{\mcitedefaultendpunct}{\mcitedefaultseppunct}\relax
\EndOfBibitem
\bibitem[Azpiroz \latin{et~al.}(2015)Azpiroz, Mosconi, Bisquert, and
  Angelis]{azpirozDefectsMigrationMethylammonium}
Azpiroz,~J.~M.; Mosconi,~E.; Bisquert,~J.; Angelis,~F.~D. Defect migration in
  methylammonium lead iodide and its role in perovskite solar cell operation.
  \emph{Energy Environ. Sci.} \textbf{2015}, \emph{8}, 2118--2127\relax
\mciteBstWouldAddEndPuncttrue
\mciteSetBstMidEndSepPunct{\mcitedefaultmidpunct}
{\mcitedefaultendpunct}{\mcitedefaultseppunct}\relax
\EndOfBibitem
\bibitem[Eames \latin{et~al.}(2015)Eames, Frost, Barnes, O'Regan, Walsh, and
  Islam]{eamesIonicTransportHybrid2015}
Eames,~C.; Frost,~J.~M.; Barnes,~P. R.~F.; O'Regan,~B.~C.; Walsh,~A.;
  Islam,~M.~S. Ionic Transport in Hybrid Lead Iodide Perovskite Solar Cells.
  \emph{Nat. Commun.} \textbf{2015}, \emph{6}, 7497\relax
\mciteBstWouldAddEndPuncttrue
\mciteSetBstMidEndSepPunct{\mcitedefaultmidpunct}
{\mcitedefaultendpunct}{\mcitedefaultseppunct}\relax
\EndOfBibitem
\bibitem[Mosconi \latin{et~al.}(2016)Mosconi, Meggiolaro, Snaith, Stranks, and
  Angelis]{mosconiLightinducedAnnihilationFrenkel2016}
Mosconi,~E.; Meggiolaro,~D.; Snaith,~H.~J.; Stranks,~S.~D.; Angelis,~F.~D.
  Light-Induced Annihilation of {{Frenkel}} Defects in Organo-Lead Halide
  Perovskites. \emph{Energy Environ. Sci.} \textbf{2016}, \emph{9},
  3180--3187\relax
\mciteBstWouldAddEndPuncttrue
\mciteSetBstMidEndSepPunct{\mcitedefaultmidpunct}
{\mcitedefaultendpunct}{\mcitedefaultseppunct}\relax
\EndOfBibitem
\bibitem[Meggiolaro \latin{et~al.}(2019)Meggiolaro, Mosconi, and
  De~Angelis]{meggiolaroFormationSurfaceDefects2019}
Meggiolaro,~D.; Mosconi,~E.; De~Angelis,~F. Formation of {{Surface Defects
  Dominates Ion Migration}} in {{Lead-Halide Perovskites}}. \emph{ACS Energy
  Lett.} \textbf{2019}, \emph{4}, 779--785\relax
\mciteBstWouldAddEndPuncttrue
\mciteSetBstMidEndSepPunct{\mcitedefaultmidpunct}
{\mcitedefaultendpunct}{\mcitedefaultseppunct}\relax
\EndOfBibitem
\bibitem[Du(2015)]{duDensityFunctionalCalculations2015}
Du,~M.-H. Density {{Functional Calculations}} of {{Native Defects}} in
  {{CH3NH3PbI3}}: {{Effects}} of {{Spin}}\textendash{{Orbit Coupling}} and
  {{Self-Interaction Error}}. \emph{J. Phys. Chem. Lett.} \textbf{2015},
  \emph{6}, 1461--1466\relax
\mciteBstWouldAddEndPuncttrue
\mciteSetBstMidEndSepPunct{\mcitedefaultmidpunct}
{\mcitedefaultendpunct}{\mcitedefaultseppunct}\relax
\EndOfBibitem
\bibitem[Meggiolaro \latin{et~al.}(2018)Meggiolaro, Motti, Mosconi, Barker,
  Ball, Perini, Deschler, Petrozza, and
  Angelis]{meggiolaroIodineChemistryDetermines2018}
Meggiolaro,~D.; Motti,~S.~G.; Mosconi,~E.; Barker,~A.~J.; Ball,~J.; Perini,~C.
  A.~R.; Deschler,~F.; Petrozza,~A.; Angelis,~F.~D. Iodine Chemistry Determines
  the Defect Tolerance of Lead-Halide Perovskites. \emph{Energy Environ. Sci.}
  \textbf{2018}, \emph{11}, 702--713\relax
\mciteBstWouldAddEndPuncttrue
\mciteSetBstMidEndSepPunct{\mcitedefaultmidpunct}
{\mcitedefaultendpunct}{\mcitedefaultseppunct}\relax
\EndOfBibitem
\bibitem[Chen \latin{et~al.}(2019)Chen, Wang, and Beard]{Ultrafastprobes}
Chen,~X.; Wang,~K.; Beard,~M.~C. Ultrafast probes at the interfaces of solar
  energy conversion materials. \emph{Phys. Chem. Chem. Phys.} \textbf{2019},
  \emph{21}, 16399--16407\relax
\mciteBstWouldAddEndPuncttrue
\mciteSetBstMidEndSepPunct{\mcitedefaultmidpunct}
{\mcitedefaultendpunct}{\mcitedefaultseppunct}\relax
\EndOfBibitem
\bibitem[Yang \latin{et~al.}(2017)Yang, Yang, Moore, Miller, Zhu, and
  Beard]{Topandbottom}
Yang,~Y.; Yang,~M.; Moore,~D.~T.; Miller,~E.~M.; Zhu,~K.; Beard,~M.~C. Top and
  bottom surfaces limit carrier lifetime in lead iodide perovskite films.
  \emph{Nat. Energy} \textbf{2017}, \emph{2}, 16207\relax
\mciteBstWouldAddEndPuncttrue
\mciteSetBstMidEndSepPunct{\mcitedefaultmidpunct}
{\mcitedefaultendpunct}{\mcitedefaultseppunct}\relax
\EndOfBibitem
\bibitem[Stranks \latin{et~al.}(2013)Stranks, Eperon, Grancini, Menelaou,
  Alcocer, Leijtens, Herz, Petrozza, and Snaith]{SnaithTRPL}
Stranks,~S.~D.; Eperon,~G.~E.; Grancini,~G.; Menelaou,~C.; Alcocer,~M. J.~P.;
  Leijtens,~T.; Herz,~L.~M.; Petrozza,~A.; Snaith,~H.~J. Electron-Hole
  Diffusion Lengths Exceeding 1 Micrometer in an Organometal Trihalide
  Perovskite Absorber. \emph{Science} \textbf{2013}, \emph{342}, 341--344\relax
\mciteBstWouldAddEndPuncttrue
\mciteSetBstMidEndSepPunct{\mcitedefaultmidpunct}
{\mcitedefaultendpunct}{\mcitedefaultseppunct}\relax
\EndOfBibitem
\bibitem[Zhai \latin{et~al.}(2020)Zhai, Wang, Zhang, Xiao, Rose, Zhu, and
  Beard]{IndividualMobilities}
Zhai,~Y.; Wang,~K.; Zhang,~F.; Xiao,~C.; Rose,~A.~H.; Zhu,~K.; Beard,~M.~C.
  Individual Electron and Hole Mobilities in Lead-Halide Perovskites Revealed
  by Noncontact Methods. \emph{ACS Energy Lett.} \textbf{2020}, \emph{5},
  47--55\relax
\mciteBstWouldAddEndPuncttrue
\mciteSetBstMidEndSepPunct{\mcitedefaultmidpunct}
{\mcitedefaultendpunct}{\mcitedefaultseppunct}\relax
\EndOfBibitem
\bibitem[Shi \latin{et~al.}(2018)Shi, Yao, Hou, Guo, Li, Wei, Ding, Li, Zhao,
  and Zhang]{UnravelingShi}
Shi,~B.; Yao,~X.; Hou,~F.; Guo,~S.; Li,~Y.; Wei,~C.; Ding,~Y.; Li,~Y.;
  Zhao,~Y.; Zhang,~X. Unraveling the Passivation Process of PbI2 to Enhance the
  Efficiency of Planar Perovskite Solar Cells. \emph{J. Phys. Chem. C}
  \textbf{2018}, \emph{122}, 21269--21276\relax
\mciteBstWouldAddEndPuncttrue
\mciteSetBstMidEndSepPunct{\mcitedefaultmidpunct}
{\mcitedefaultendpunct}{\mcitedefaultseppunct}\relax
\EndOfBibitem
\bibitem[Zhang \latin{et~al.}(2020)Zhang, Turiansky, Shen, and {Van de
  Walle}]{zhangIodineInterstitialsCause2020}
Zhang,~X.; Turiansky,~M.~E.; Shen,~J.-X.; {Van de Walle},~C.~G. Iodine
  Interstitials as a Cause of Nonradiative Recombination in Hybrid Perovskites.
  \emph{Phys. Rev. B} \textbf{2020}, \emph{101}, 140101\relax
\mciteBstWouldAddEndPuncttrue
\mciteSetBstMidEndSepPunct{\mcitedefaultmidpunct}
{\mcitedefaultendpunct}{\mcitedefaultseppunct}\relax
\EndOfBibitem
\bibitem[Huang \latin{et~al.}(2021)Huang, Vardeny, Wang, Ahmad, Chanana,
  Vetter, Yang, Liu, Galli, Amassian, Vardeny, and
  Sun]{huangObservationSpatiallyResolved2021}
Huang,~Z.; Vardeny,~S.~R.; Wang,~T.; Ahmad,~Z.; Chanana,~A.; Vetter,~E.;
  Yang,~S.; Liu,~X.; Galli,~G.; Amassian,~A.; Vardeny,~Z.~V.; Sun,~D.
  Observation of Spatially Resolved {{Rashba}} States on the Surface of
  {{CH3NH3PbBr3}} Single Crystals. \emph{Appl. Phys. Rev.} \textbf{2021},
  \emph{8}, 031408\relax
\mciteBstWouldAddEndPuncttrue
\mciteSetBstMidEndSepPunct{\mcitedefaultmidpunct}
{\mcitedefaultendpunct}{\mcitedefaultseppunct}\relax
\EndOfBibitem
\bibitem[Myung \latin{et~al.}(2018)Myung, Javaid, Kim, and
  Lee]{myungRashbaDresselhausEffect2018}
Myung,~C.~W.; Javaid,~S.; Kim,~K.~S.; Lee,~G. Rashba\textendash{{Dresselhaus
  Effect}} in {{Inorganic}}/{{Organic Lead Iodide Perovskite Interfaces}}.
  \emph{ACS Energy Lett.} \textbf{2018}, \emph{3}, 1294--1300\relax
\mciteBstWouldAddEndPuncttrue
\mciteSetBstMidEndSepPunct{\mcitedefaultmidpunct}
{\mcitedefaultendpunct}{\mcitedefaultseppunct}\relax
\EndOfBibitem
\bibitem[Motti \latin{et~al.}(2019)Motti, Meggiolaro, Barker, Mosconi, Perini,
  Ball, Gandini, Kim, De~Angelis, and
  Petrozza]{mottiControllingCompetingPhotochemical2019}
Motti,~S.~G.; Meggiolaro,~D.; Barker,~A.~J.; Mosconi,~E.; Perini,~C. A.~R.;
  Ball,~J.~M.; Gandini,~M.; Kim,~M.; De~Angelis,~F.; Petrozza,~A. Controlling
  Competing Photochemical Reactions Stabilizes Perovskite Solar Cells.
  \emph{Nat. Photonics} \textbf{2019}, \emph{13}, 532--539\relax
\mciteBstWouldAddEndPuncttrue
\mciteSetBstMidEndSepPunct{\mcitedefaultmidpunct}
{\mcitedefaultendpunct}{\mcitedefaultseppunct}\relax
\EndOfBibitem
\bibitem[Yang \latin{et~al.}(2019)Yang, Chen, Mosconi, Fang, Xiao, Wang, Zhou,
  Yu, Zhao, Gao, Angelis, and Huang]{yangStabilizingHalidePerovskite2019}
Yang,~S.; Chen,~S.; Mosconi,~E.; Fang,~Y.; Xiao,~X.; Wang,~C.; Zhou,~Y.;
  Yu,~Z.; Zhao,~J.; Gao,~Y.; Angelis,~F.~D.; Huang,~J. Stabilizing Halide
  Perovskite Surfaces for Solar Cell Operation with Wide-Bandgap Lead Oxysalts.
  \emph{Science} \textbf{2019}, \emph{365}, 473--478\relax
\mciteBstWouldAddEndPuncttrue
\mciteSetBstMidEndSepPunct{\mcitedefaultmidpunct}
{\mcitedefaultendpunct}{\mcitedefaultseppunct}\relax
\EndOfBibitem
\bibitem[Herz(2017)]{HerzChargeCarrierPerspective}
Herz,~L.~M. Charge-Carrier Mobilities in Metal Halide Perovskites: Fundamental
  Mechanisms and Limits. \emph{ACS Energy Lett.} \textbf{2017}, \emph{2},
  1539--1548\relax
\mciteBstWouldAddEndPuncttrue
\mciteSetBstMidEndSepPunct{\mcitedefaultmidpunct}
{\mcitedefaultendpunct}{\mcitedefaultseppunct}\relax
\EndOfBibitem
\end{mcitethebibliography}


\section*{Supplementary material}

# Computational Details

All DFT calculations have been performed using the plane wave code Quantum Espresso.[1,2] The Perdew–Burke-Ernzerhof (PBE) exchange-correlation functional[3] without spin-orbit coupling was used for geometry optimizations. On the relaxed structures, single point DFT calculations using the Heyd–Scuseria–Ernzerhof (HSE) hybrid functional[4] with spin-orbit coupling were performed. The DFT-D3 scheme proposed by Grimme was used to account for the van der Waals dispersion.[5] For PBE calculations, we used the ultrasoft GBRV pseudopotentials with the suggested energy cutoffs of 40 Ry for the wavefunction and 320 Ry for the density.[6] For the HSE calculations, we used SG15 ONCV pseudopotentials[7] with a 70 Ry energy cutoff for the wavefunction, 280 Ry for the density, and 70 Ry for the exact exchange operator. For interfacial calculations, we used I pseudopotential with 7 valence electrons instead of 17 to reduce the computational cost. The fraction of the exact exchange used for both MAPI and $PbI_2$ was 0.43 which has been shown to provide a better agreement with the experimental band gap of MAPI.[8] For $PbI_2$ as well, the band gap using this value of exact exchange is in agreement with experiment as shown in Table S1.[9] For surfaces and interfaces, the periodic images were separated by 15 Å of vacuum for PBE calculations and 10 Å for the HSE calculations to save memory. HSE calculations performed using 15 Å vacuum showed negligible change in the energies. The pymatgen[10] and atomic simulation environments[11] were used for generating surface and interfacial structures while the visualization was done using the VESTA package.[12] For defect formation energy calculations using HSE functional, reduced energy cutoffs of 40 Ry for the wavefunction and 80 Ry for the exact exchange operator were used which have been shown to agree well with those calculated at 70 Ry energy cutoffs due to error cancellation for the pristine and defective structures.[13] For MAPI, we

Table S1: Band gaps obtained using hybrid functional with fraction of exact exchange = 0.43.

| Material | Band gap (eV) |
|---|---|
| MAPI | 1.42 |
| $PbI_2$ | 2.36 |

used the experimental lattice constants 8.849 Å and 12.642 Å .[14] For $PbI_2$, the structure # 9009114[15] from crystallography open database[16] with spacegroup P$\bar{3}$m1 was used. The ionic positions are relaxed according to convergence criteria of at most 0.001 Ry/Bohr for forces and 0.0001 Ry for energies. For some pristine structures, reduced force convergence criteria had to be used to obtain the defect formation energies due to the shallow nature of the energy landscape. An example of an interfacial defect is shown in Fig. S1

The interfaces identified are shown in Table S2 have in plane lattice vectors normal to each other. This enforces the $PbI_2$ surface to be $(1\,0\,\bar{1}\,0)$ with its lattice vectors b and c in the plane of the interface.

Table S2: Possible interfaces between different surfaces of MAPI and $PbI_2$. All interfaces studied use the $PbI_2$ termination of MAPI in order to study $PbI_2$ excess effects. The table shows the surface facets of MAPI and $PbI_2$, the in plane lattice vectors $a_1$ and $a_2$ and strains in those directions. For interfaces besides those with MAPI (0 0 1) in rows 1 and 2, we fixed the in-plane lattice vectors to the MAPI values since the existence of small amounts of $PbI_2$ is not expected to influence MAPI axes considerably.

| Lattice vector | MAPI | $PbI_2$ | Interface | strain MAPI (%) | strain $PbI_2$ (%) |
|---|---|---|---|---|---|
| \multicolumn{6}{c}{MAPI (0 0 1) with Pb terminated $PbI_2$ (1 0 $\bar{1}$ 0), with cell relaxation} | | | | | |
| $a_1$ | a=8.849 | 2a=9.11 | 9.06 | 2.33 | -0.60 |
| $a_2$ | a=8.849 | c=6.977 | 8.38 | -5.33 | 20.07 |
| \multicolumn{6}{c}{MAPI (0 0 1) with I terminated $PbI_2$ (1 0 $\bar{1}$ 0), with cell relaxation} | | | | | |
| $a_1$ | a=8.849 | 2a=9.11 | 9.070 | 2.50 | -0.44 |
| $a_2$ | a=8.849 | c=6.977 | 8.358 | -5.54 | 19.80 |
| \multicolumn{6}{c}{MAPI (1 1 0) with $PbI_2$ (1 0 $\bar{1}$ 0); Lattice vectors fixed to MAPI values; two Interfaces: Pb and I terminations of $PbI_2$} | | | | | |
| $a_1$ | c=12.642 | 3b=13.665 | 12.642 | 0 | -7.49 |
| $a_2$ | $\sqrt{2}$a=12.514 | 2c=13.954 | 12.514 | 0 | -10.32 |
| \multicolumn{6}{c}{MAPI (0 0 1) rotated 45 degrees about [0 0 1] axis with $PbI_2$ (1 0 $\bar{1}$ 0); Lattice vectors fixed to MAPI values; two Interfaces: Pb and I terminations of $PbI_2$} | | | | | |
| $a_1$ | $\sqrt{2}$b=12.514 | 3b=13.665 | 12.514 | 0 | -8.42 |
| $a_2$ | $\sqrt{2}$a=12.514 | 2c=13.954 | 12.514 | 0 | -10.32 |

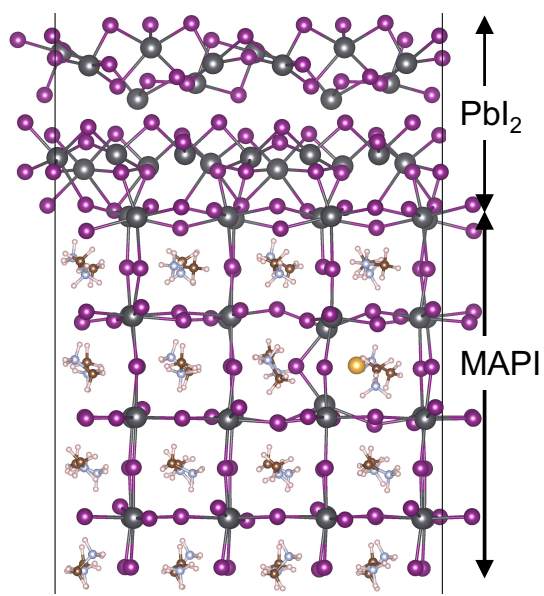

Figure S1: An $I^-/V_I^+$ Frenkel pair in MAPI located close to the interface with $PbI_2$. The interstitial I is shown in yellow and the vacancy is located in the layer above it. The interface is of the last type mentioned in the table i.e., MAPI $(0\,0\,1)$ rotated 45 degrees about $[0\,0\,1]$ axis with $PbI_2$ $(1\,0\,\bar{1}\,0)$

# Experimental Details

**Sample preparation:**

MAPI films were prepared according to previous reports.[17] 0.5 x 1 cm quartz substrates were prepared by sonicating in isopropyl alcohol followed by sonicating in acetone. The quartz slides were ozone cleaned for 15 minutes immediately before perovskite deposition. MAPI precursor solutions were made by dissolving MAI (Dyesol) and $PbI_2$ (Alfa Aesar, 99.9985%, metal basis, power) in a 9:1 N,N-dimethylformamide (DMF, Sigma Aldrich, anhydrous, 99.8%):dimethylsulfoxide (DMSO, Sigma Aldrich, anhydrous, >99.9%) solution. For the stoichiometric films (0% excess PbI2), 1.5 M films were made. To create films of other stoichiometric values, 2.5%, 5%, and 10% excess $PbI_2$ by molarity was added. These films were used for TR, TRPL, and XRD measurements.

Additional films for TRPL measurements were made with 0.75 and 0.375 M solutions. All solutions were prepared in a glovebox. Films were deposited using an antisolvent spin-coating method. 50 $\mu$L of the perovskite precursor solution was statically drop cast onto the films. The films were then spun at 4000 rpms for 25 seconds, with 60 uL of chlorobenzene (Sigma Aldrich, anhydrous, >99.8%) dynamically spun cast onto the film after 10 seconds of spin coating. The films were annealed for 1 minute at 65 C followed by 2 minutes at 100 C. Films were kept in a nitrogen flow box after deposition to ensure the films did not degrade under ambient conditions.

**Transient reflectance measurements:**

Transient reflectance measurements were performed on a Coherent Laser (800 nm fundamental beam, 1 kHz rep rate, 3 mJ/pulse, 100 fs pulse width) with a Helios Ultrafast spectrometer used for detection. The fundamental beam was split into the pump and probe pulse directly after the Coherent laser. The pump laser was tuned with a Palitra Duo OPA to obtain varying excitation wavelengths needed for TR measurements. The probe beam was sent through a delay line before being transformed into a white light continuum using a Sapphire crystal. The pump and probe pulses were aligned on the sample, with the probe

pulse at a 45° angle to the sample. All samples were measured under ambient conditions. The excitation density for each wavelength was kept so that $N_0$ was approximately $4 \times 10^{17}$ charge carriers/cm$^3$.

**Time resolved photoluminescence:**

A Hamamatsu C10910-04 streak camera was used for detection and an NKT supercontinuum fiber laser (SuperK EXU-6-PP) for excitation. Samples were excited with 600 nm light.

**Profilometry:**

A Dektak 8 Advanced Development Profilometer was used to obtain sample thickness information. At least 3 scans were performed on each film and averaged to find the film thickness.

**XRD:**

X-ray diffraction patterns were measured with a Bruker D8 diffractometer equipped with a Hi-Star 2D area detector and a Cu K$\alpha$ radiation source. Scherrer analysis was performed by fitting peaks to a Lorentzian profile.

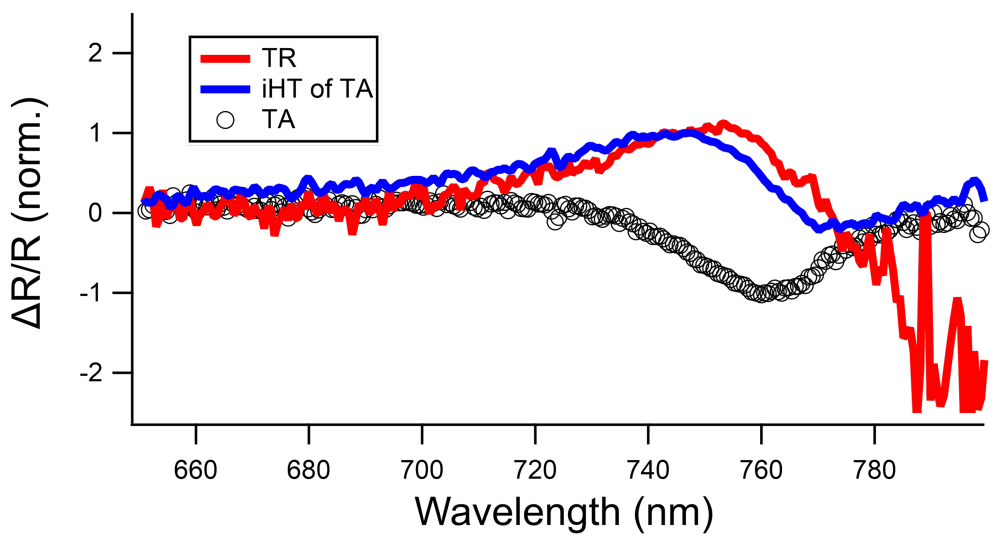

Figure S2: Transient absorption (grey circles), transient reflectance (red trace), and the inverse Hilbert transform of the transient absorption spectrum (blue trace). The TA signal shows a ground state bleach around 760 nm, typical of MAPI. Using the Kramers-Kronig relation, the transient absorption spectrum (grey) can be transformed into a simulation of the transient reflectance spectrum (blue). At wavelengths smaller than 750 nm (higher energy), the transform of the TA (blue) matches exactly with the transient reflectance spectrum (red). This indicates that at wavelength where the inverse Hilbert transform of the TA matches the TR, only changes in reflectance at the front (surface) of the sample is probed. Below 760 nm, an interference regime where signal from both the front and back surfaces of the perovskite is probed. This indicates that the kinetics for analysis should be taken from higher than 760 nm.

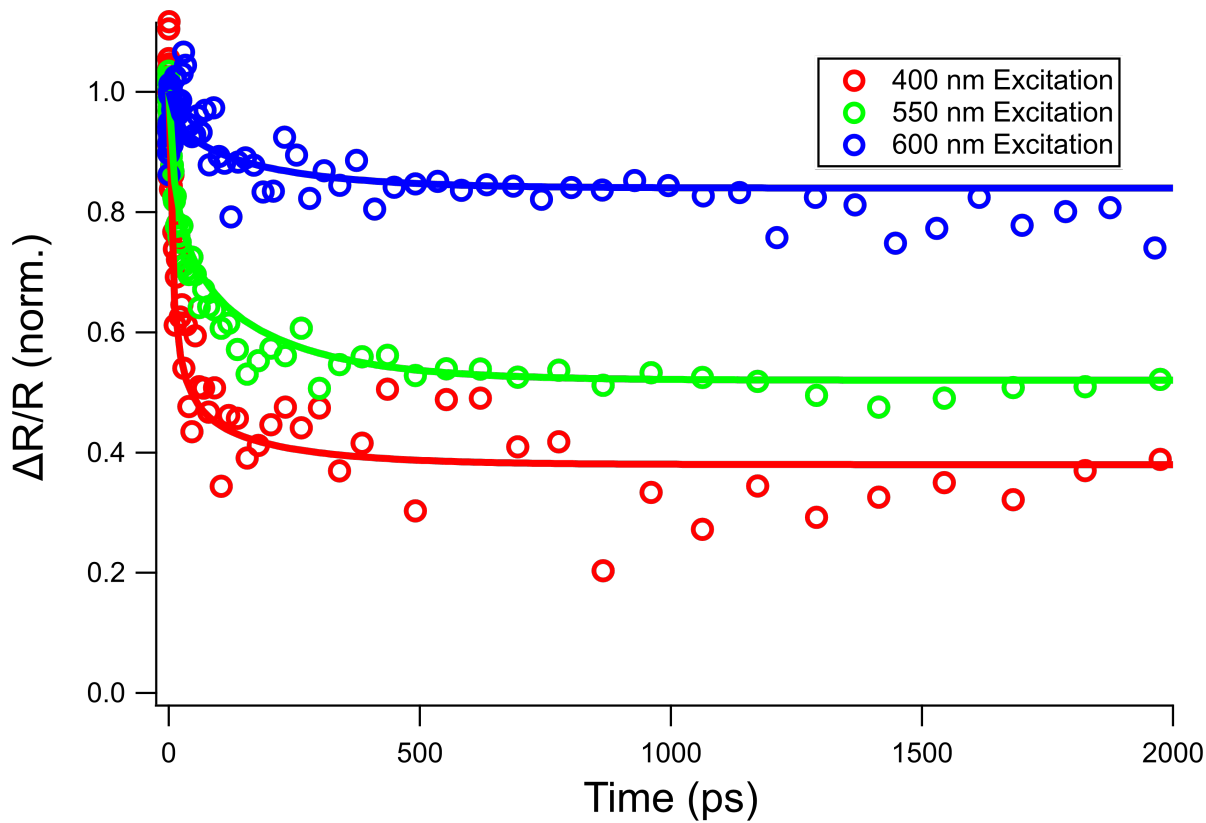

Figure S3: Transient reflectance kinetics (taken at 700 nm) for the 10% excess sample pumped at different wavelengths. Since the absorption coefficient of MAPI increases with increasing pump energies, lower energy pumps are able to excite farther into the film compared to high energy pumps. This allows for larger volumes of the film, and therefore more bulk-like (vs. surface) behavior, with decreasing pump energy. The higher the pump energy, the shorter the kinetic lifetime. Since the change in the signal across all three wavelengths is primarily concentrated to the first few hundred ps, then diffusion can be identified as the mechanism that contributes to the change in signal. Therefore, diffusion dominates at the surface. This data can be fit to a global analysis that can quantitatively extract the diffusion coefficient.[18] The fitting procedure was performed on raw data without normalization. The 10% excess sample is representative for the data for 0, 2.5, and 5% excess as well.

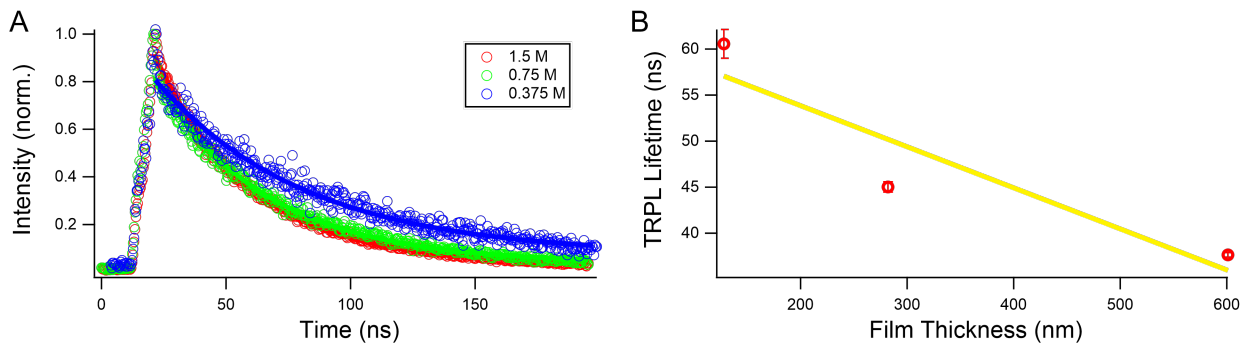

Figure S4: (A) Time resolved photoluminescence of 0% excess films. Films were made with precursors with varying molarity, which yields films with different thicknesses (as measured by profilometry, B). The films were excited with 600 nm light and the PL lifetime was then fit to a monoexponential decay. (B) The extracted lifetimes were then plotted against the film thickness so that a linear fit between TRPL lifetime and film thickness could be found. The inverse of this relationship yields the surface recombination lifetime in m/s. These results are reported in Fig. 4 of the main text.

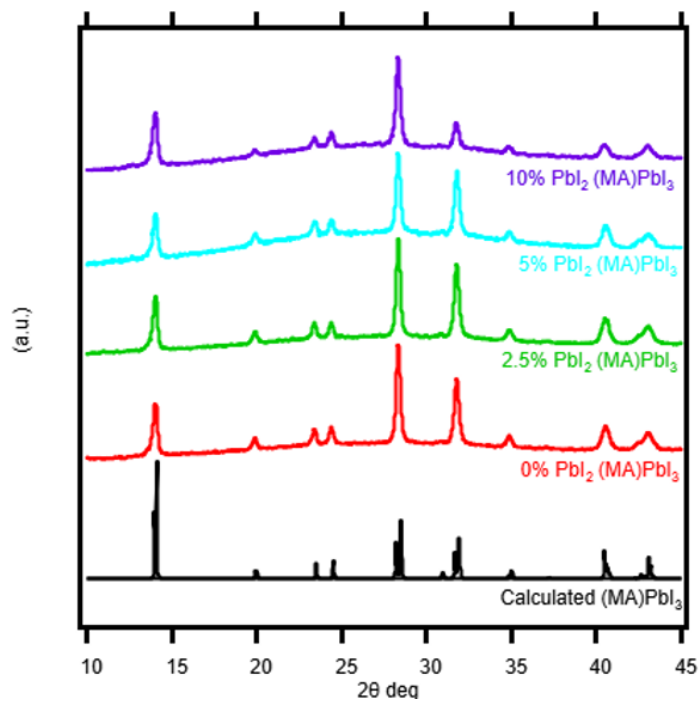

Figure S5: X-ray diffraction (XRD) patterns of freshly prepared MAPI thin films with varying $PbI_2$ content. The XRD patterns match well to the calculated MAPI pattern[19] and are absent of any impurity peaks. The average crystallite size calculated from the Scherrer equation is 56 nm, 56 nm, 51 nm, 52 nm for 0%, 2.5%, 5%, 10% $PbI_2$, respectively, indicating there is little variation with perovskite grain size with varying ratios of $PbI_2$ in the precursor.

# References

(1) Giannozzi, P. et al. QUANTUM ESPRESSO: a modular and open-source software project for quantum simulations of materials. *J. Phys.: Condens. Matter* **2009**, *21*, 395502.

(2) Giannozzi, P. et al. Advanced capabilities for materials modelling with Quantum ESPRESSO. *J. Phys.: Condens. Matter* **2017**, *29*, 465901.

(3) Perdew, J. P.; Burke, K.; Ernzerhof, M. Generalized Gradient Approximation Made Simple. *Phys. Rev. Lett.* **1996**, *77*, 3865–3868.

(4) Heyd, J.; Scuseria, G. E.; Ernzerhof, M. Hybrid functionals based on a screened Coulomb potential. *J. Chem. Phys.* **2003**, *118*, 8207–8215.

(5) Grimme, S.; Antony, J.; Ehrlich, S.; Krieg, H. A consistent and accurate ab initio parametrization of density functional dispersion correction (DFT-D) for the 94 elements H-Pu. *J. Chem. Phys.* **2010**, *132*, 154104.

(6) Garrity, K. F.; Bennett, J. W.; Rabe, K. M.; Vanderbilt, D. Pseudopotentials for high-throughput DFT calculations. *Comput. Mater. Sci.* **2014**, *81*, 446–452.

(7) Schlipf, M.; Gygi, F. Optimization algorithm for the generation of ONCV pseudopotentials. *Comput. Phys. Commun.* **2015**, *196*, 36–44.

(8) Du, M.-H. Density Functional Calculations of Native Defects in CH3NH3PbI3: Effects of Spin–Orbit Coupling and Self-Interaction Error. *J. Phys. Chem. Lett.* **2015**, *6*, 1461–1466.

(9) Supasai, T.; Rujisamphan, N.; Ullrich, K.; Chemseddine, A.; Dittrich, T. Formation of a Passivating CH3NH3PbI3/PbI2 Interface during Moderate Heating of CH3NH3PbI3 Layers. *Appl. Phys. Lett.* **2013**, *103*, 183906.

(10) Ong, S. P.; Richards, W. D.; Jain, A.; Hautier, G.; Kocher, M.; Cholia, S.; Gunter, D.; Chevrier, V. L.; Persson, K. A.; Ceder, G. Python Materials Genomics (pymatgen): A robust, open-source python library for materials analysis. *Comput. Mater. Sci.* **2013**, *68*, 314 – 319.

(11) Larsen, A. H. et al. The atomic simulation environment—a Python library for working with atoms. *J. Phys.: Condens. Matter* **2017**, *29*, 273002.

(12) Momma, K.; Izumi, F. VESTA 3 for three-dimensional visualization of crystal, volumetric and morphology data. *J. Appl. Crystallogr.* **2011**, *44*, 1272–1276.

(13) Meggiolaro, D.; De Angelis, F. First-Principles Modeling of Defects in Lead Halide Perovskites: Best Practices and Open Issues. *ACS Energy Lett.* **2018**, *3*, 2206–2222.

(14) Stoumpos, C. C.; Malliakas, C. D.; Kanatzidis, M. G. Semiconducting Tin and Lead Iodide Perovskites with Organic Cations: Phase Transitions, High Mobilities, and Near-Infrared Photoluminescent Properties. *Inorg. Chem.* **2013**, *52*, 9019–9038.

(15) Wyckoff, R. W. G. Crystal structures, 2nd edition. *Acta Crystallogr.* **1963**, *1*, 239–444.

(16) Gražulis, S.; Chateigner, D.; Downs, R. T.; Yokochi, A. F. T.; Quirós, M.; Lutterotti, L.; Manakova, E.; Butkus, J.; Moeck, P.; Bail, A. L. Crystallography Open Database – an open-access collection of crystal structures. *J. Appl. Cryst.* **2009**, *42*, 726–729.

(17) Elmelund, T.; Scheidt, R. A.; Seger, B.; Kamat, P. V. Bidirectional Halide Ion Exchange in Paired Lead Halide Perovskite Films with Thermal Activation. *ACS Energy Lett.* **2019**, *4*, 1961–1969.

(18) Zhai, Y.; Wang, K.; Zhang, F.; Xiao, C.; Rose, A. H.; Zhu, K.; Beard, M. C. Individual Electron and Hole Mobilities in Lead-Halide Perovskites Revealed by Noncontact Methods. *ACS Energy Lett.* **2020**, *5*, 47–55.

(19) Szafrański, M.; Katrusiak, A. Mechanism of Pressure-Induced Phase Transitions, Amorphization, and Absorption-Edge Shift in Photovoltaic Methylammonium Lead Iodide. *J. Phys. Chem. Lett.* **2016**, *7*, 3458–3466, PMID: 27538989.

# Graphical TOC Entry

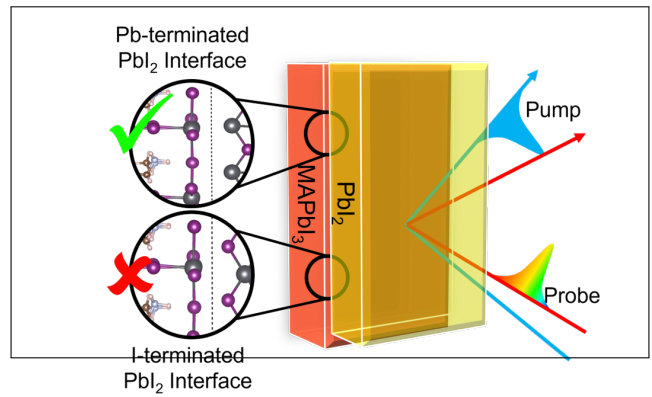


\end{document}